\documentclass[aps,nofootinbib,showpacs,preprintnumbers,amsmath,amssymb]{revtex4}
\def\be{\begin{equation}}
\def\ee{\end{equation}}
\def\ba{\begin{eqnarray}}
\def\ea{\end{eqnarray}}
\def\bea{\begin{eqnarray}}
\def\eea{\end{eqnarray}}
\def\bes{\begin{subequations}}
\def\ees{\end{subequations}}
\newcommand{\A}{{\mathcal{A}}}

\newcommand{\MSbar}{\overline{\rm MS}}  
\usepackage{graphics}
\usepackage{graphicx}
\usepackage{dcolumn}
\usepackage{bm}
\usepackage{epsfig}
\usepackage{graphicx,color}
\usepackage{multirow}
\usepackage{xcolor}

\begin{document}


\title{pQCD running couplings finite and monotonic in the infrared: when do they reflect the holomorphic properties of spacelike observables?} 
\author{Carlos Contreras$^1$}
\author{Gorazd Cveti\v{c}$^1$}
\author{Oscar Orellana$^2$}

\affiliation{$^1$Department of Physics, Universidad T{\'e}cnica Federico Santa Mar{\'\i}a, Casilla 110-V, Valpara{\'\i}so, Chile\\
$^2$Department of Mathematics, Universidad T\'ecnica Federico Santa Mar\'ia, Casilla 110-V, Valpara\'iso, Chile}

\date{\today}

\begin{abstract}
  We investigate a large class of perturbative QCD (pQCD) renormalization schemes whose beta functions $\beta(a)$ are meromorphic functions of the running coupling and give finite positive value of the coupling $a(Q^2)$ in the infrared regime (``freezing''), $a(Q^2) \to a_0$ for $Q^2 \to 0$.  Such couplings automatically have no singularities on the positive axis of the squared momenta $Q^2$ ($ \equiv -q^2$). Explicit integration of the renormalization group equation (RGE) leads to the implicit (inverted) solution for the coupling, of the form $\ln (Q^2/Q^2_{\rm in}) = {\cal H}(a)$. An analysis of this solution leads us to an algebraic algorithm for the search of the Landau singularities of $a(Q^2)$ on the first Riemann sheet of the complex $Q^2$-plane, i.e., poles and branching points (with cuts) outside the negative semiaxis. We present specific representative examples of the use of such algorithm, and compare the found Landau singularities with those seen after the 2-dimensional numerical integration of the RGE in the entire first Riemann sheet, where the latter approach is numerically demanding and may not always be precise. The specific examples suggest that the presented algebraic approach is useful to find out whether the running pQCD coupling has Landau singularities and, if yes, where precisely these singularities are.
\end{abstract}

\maketitle

\section{Introduction}
\label{sec:intr}

According to general principles of Quantum Field Theories, the physical spacelike observables ${\cal D}(Q^2)$ (such as the quark current correlators) and even unphysical amplitudes (such as the dressing functions of quark and transverse gluon propagators in QCD) are holomorphic (analytic) functions in the complex $Q^2$-plane (where $Q^2 \equiv - q^2 = - (q^0)^2 + {\vec q}^2$) except on the negative $Q^2$ semiaxis \cite{Oehme,BS}. On the other hand, QCD running coupling $a(Q^2) \equiv \alpha_s(Q^2)/\pi$ can be defined, in a specific renormalization scheme, as a product of the Landau gauge gluon dressing function and the square of the ghost dressing function \cite{LerchSme}. Further, the leading-twist part of the spacelike physical QCD amplitudes ${\cal D}(Q^2)$ is a function of the running coupling, ${\cal D}(Q^2)^{\rm (l.t.)} = {\cal F}(a(\kappa Q^2); \kappa)$ (where $\kappa \sim 1$ is a positive renormalization scale parameter). Therefore, a natural consequence of the holomorphic behaviour of QCD amplitudes ${\cal D}(Q^2)$ would be that QCD running coupling $a(Q^2)$ reflected these properties, i.e., that $a(Q^2)$ were a holomorphic function in the complex $Q^2$-plane with the exception of a negative semiaxis, $Q^2 \in \mathbb{C} \backslash (-\infty, -M_{\rm thr}^2]$, where $M_{\rm thr}$ is a threshold mass, $0 \leq M_{\rm thr}^2 \lesssim 0.1 \ {\rm GeV}^2$.

  However, the QCD coupling $a(Q^2)$ \textcolor{black}{is evaluated often in such renormalization schemes in which it} is not an observable, and consequently $a(Q^2)$ is not a holomorphic [on $\mathbb{C} \backslash (-\infty,0)$] function, but it may have singularities in the mentioned region, called Landau singularities. \textcolor{black}{These singularities are a serious problem especially in evaluations of low-energy QCD observables, where the coupling often has to be evaluated in the regimes of the complex $Q^2$-plane which are close to those singularities and thus the obtained values lose predictability.} Therefore, it is important to have a reliable method to find whether such Landau singularities exist, and if they exist, where in the complex $Q^2$-plane they are situated and what is their nature.

\textcolor{black}{Specifically, if the considered observable ${\cal D}(Q^2)$ is spacelike and the spacelike momentum $Q^2$ is positive, the leading-twist part of ${\cal D}(Q^2)$ is evaluated as a perturbation series in powers of $a(\kappa Q^2)$ where $\kappa$ is a positive renormalization scale parameter ($\kappa \sim 1$); if the coupling $a(Q^{'2})$ has Landau singularities in the complex $Q^{'2}$-plane at values on or close to the positive semiaxis, then the evaluation of  ${\cal D}(Q^2)$ becomes unreliable for $Q^2$ close to such Landau singularities.}

\textcolor{black}{Further, if the considered QCD observable ${\cal R}(s)$ is timelike ($s=-Q^2>0$), then it is usually evaluated as a contour integral involving the corresponding spacelike quantity $\Pi (Q^2)$ in the complex $Q^{'2}$-plane, with a contour of radius $|Q^{'2}| \sim s$. In such a case, there are at least two problems appearing when $a(Q^{'2})$ has Landau singularities in the complex $Q^{'2}$-plane. The first is the following: the quantity ${\cal R}(s)$ is originally expressed as an integral involving the corresponding physical (measured) spectral function $\omega(\sigma) = {\rm Im} \Pi(-\sigma - i \epsilon)$ along the physical cut $0 < \sigma < s$; this integral cannot be evaluated directly in pQCD; it is transformed via the Cauchy theorem into a contour integral involving  $\Pi (Q^{'2})$ along a circle of radius $|Q^{'2}| \sim s$ (a form of sum rules). In pQCD, the leading-twist part of the spacelike quantity $\Pi (Q^{'2})$ in this contour integral is usually expressed as a perturbation series in powers of $a(\kappa Q^{'2})$ where $\kappa$ is a positive renormalization scale parameter, $\kappa \sim 1$. If the pQCD coupling has Landau singularities, the evaluated  $\Pi (Q^{'2}) = {\cal F}(a(\kappa Q^{'2}))$ function does not possess the holomorphic properties in the complex $Q^{'2}$-plane (outside the negative axis) which it was assumed to possess when applying the Cauchy theorem. The mentioned sum rule relation is thus inconsistent in the case of pQCD with Landau singularities. The second problem that can appear here is more of a practical nature: if there are Landau singularities $Q_{*}^2$ in the complex $Q^{'2}$-plane such that $|Q_{*}^2| \sim s$, then the contour integral may come close to such singularities and the evaluation may turn numerically unstable.}

The perturbative QCD (pQCD) frameworks usually used in the literature are the $\MSbar$-type mass independent renormalization schemes (such as $\MSbar$, 't Hooft, MiniMOM, Lambert schemes), which give the running coupling $a(Q^2)$ which is not holomorphic in the mentioned sense, but has a (Landau) branching point at $Q_{\ast}^2 > 0$ ($\sim 0.1$-$1 \ {\rm GeV}^2$) for the cut, i.e., the cut reaches beyond the negative semiaxis to the positive IR regime, i.e., there is a Landau ghost cut $(0,Q_{\ast}^2)$. Further, the coupling often diverges at the branching point, $a(Q_{\ast}^2) = \infty$ (Landau pole). \textcolor{black}{These properties are mathematical consequences of the form of the beta function $\beta(a)$ which appears in the RGE determining the flow of $a(Q^2)$ with the squared momentum $Q^2$.} These properties contradict the earlier mentioned holomorphic properties for $a(Q^2)$ which are motivated physically. If the Landau branching point $Q_{\ast}^2$ is on the (positive) real axis, it is relatively straightforward to encounter it in practice, for example by one-dimensional numerical integration of the RGE along the positive $Q^2$-axis. On the other hand, if there are no Landau singularities on the positive real semiaxis, they could still appear within the complex plane $Q^2 \in \mathbb{C} \backslash  \mathbb{R}$; in such a case, it may be practically more difficult to find whether such singularities exist, and if they do exist, where they are and what is their nature. In this work we will concentrate on this problem, in the case of pQCD couplings in large classes of mass-independent renormalization schemes.

In our work, the considered QCD coupling $a(Q^2) \equiv \alpha_s(Q^2)/\pi$ will be such that it has so called freezing in the infrared regime, i.e., $a(0) = a_0$ is finite positive. This behaviour is suggested by the scaling solutions for the gluon and ghost propagators in the Landau gauge in the Dyson-Schwinger equations (DSE) approach \cite{LerchSme,DSEscale1,DSEscale2,DSEscale3,DSEscale4}, in the functional renormalization group (FRG) approach \cite{FRG1,FRG2,FRG3}, stochastic quantization \cite{STQ}, and by Gribov-Zwanziger approach \cite{Gr1,Gr2}.
Further, $0 < a(0) \equiv a_0 < + \infty$ is also obtained in various physically motivated models for the running QCD coupling, among them the minimal analyticity dispersive approach \cite{ShS1,ShS2,ShS3,ShS4,ShS5,ShirEPJC,APTapp1,APTapp2,APTapp3,APTapp4,APTapp5,APTapp6,APTapp7,APTapp8,BMS1,BMS2,BMS3,BMS4,APTrev1,APTrev2,APTrev3,APTrev4}
and its modifications or extensions \cite{bAPT1,bAPT2,bAPT3,bAPT4,bAPT5,bAPT6,bAPT7,bAPT8,bAPT9,bAPT10,bAPT11,bAPT12,bAPT13,bAPT14,bAPT15,bAPT16,bAPT17,bAPT18,bAPT19,bAPT20},\footnote{Similar dispersive approaches have been applied also directly to spacelike QCD amplitudes and observables \cite{dispapp1,dispapp2,dispapp3,dispapp4,dispapp5,dispapp6,dispapp7,dispapp8,dispapp9}. Dispersive approach leading to $a(0) = + \infty$ has been constructed in Refs.~\cite{Nest11,Nest12,Nest13,Nest14}.}
and the AdS/CFT correspondence modified by a dilaton backgound \cite{AdS1,AdS2}. For reviews, we refer to \cite{revGC,Brodrev}. Such a behaviour has also been suggested in \cite{DSEdecoupFreez},
where the running coupling definition involves explicitly the dynamical gluon mass and thus gives positive (nonzero) $a(0)$ even in the case of so called decoupling solution of DSE \cite{DSEdecoup1,DSEdecoup2,DSEdecoup3,DSEdecoup4,DSEdecoup5} for gluon and ghost propagator in the Landau gauge.\footnote{Some newer lattice results \cite{latt11,latt12,latt13,latt14,latt21,latt22,latt23,latt24,latt25,latt26} suggest the so called decoupling solution, i.e., that in the Landau gauge the gluon propagator is finite in the infrared and the ghost propagator is not infrared enhanced, indicating that the running coupling, if defined as the mentioned product of dressing functions, at very low positive $Q^2$ goes to zero as $\A(Q^2) \sim Q^2$. Such a behaviour of the running coupling is also suggested or obtained in the works \cite{Gribdec1,Gribdec2,Gribdec3,ArbZaits,Boucaud1,Boucaud2,mes21,mes22}. A holomorphic coupling $\A(Q^2)$ respecting this behaviour in the infrared, $\A(0)=0$, and perturbative QCD in the ultraviolet regime, has been constructed in Ref.~\cite{3dAQCD}. When defining a lattice coupling which involves the lattice-calculated 3-gluon Green function \cite{latt3gl1,latt3gl2,latt3gl3,latt3gl4}, a different but qualitatively similar behaviour [$\A_{\rm latt}(Q^2) \to 0$ when $Q^2 \to 0$] is obtained. We will not pursue these lines in this work.}
All these approaches lead to nonperturbative (NP) running coupling $\A(Q^2)$, which in general differs from the underlying perturbative coupling $a(Q^2)$ (i.e., the pQCD coupling in the same renormalization scheme) by power terms $\sim 1/(Q^2)^N$, i.e., terms of the type $\exp(- C/a)$ which cannot be Taylor-expanded around the pQCD point $a=0$.

However, there are also pure pQCD frameworks (beta functions) in which the running coupling achieves a finite positive value in the infrared limit $a(0) \equiv a_0 < \infty$. Among such couplings are those \textcolor{black}{where the coupling is a physical observable, such as in the effective (physical) charge approach \cite{ECH1,ECH2,KKP,DG} (cf.~also \cite{CKL}) where the coupling is a (spacelike or timelike) observable; such an observable can have variable and even very low momentum scales $|Q^2| < 1 \ {\rm GeV}^2$ \cite{BMM}, and such physical charges can even be related at the perturbative level to each other analytically \cite{Crewther1,Crewther2,Crewther3}.} Application of the principle of minimal sensitivity \cite{Stevenson} also leads to schemes which give finite positive value of $a(0)$. There exist yet other renormalization schemes with $a(0)>0$, namely such that the resulting pQCD coupling $a(Q^2)$ is holomorphic in $Q^2 \in \mathbb{C} \backslash (-\infty, -M_{\rm thr}^2]$ (with $0 < M_{\rm thr}^2 \sim 0.1 \ {\rm GeV}^2$) and reproduces the correct high- and low-energy QCD phenomenology \cite{anpQCD1,anpQCD2,anpQCD3}.

In Sec.~\ref{sec:impl} we define the class of considered pQCD beta functions $\beta(a)$ (i.e., renormalization schemes), which are meromorphic functions leading to a finite positive $a(0)$, present the implicit solution of the RGE in the complex $Q^2$-plane, and discuss the renormalization scheme parameters $\beta_j$ ($j \geq 2$) that such beta functions generate. In Sec.~\ref{sec:sing} we then present a practical algebraic procedure which allows us to find for a chosen beta function (in the considered class) the Landau singularities in the complex $Q^2$-plane, i.e., the points where the behaviour of the running coupling $a(Q^2)$ does not reflect the holomorphic properties of the spacelike Green functions ${\cal D}(Q^2)$ as required by the general principles of Quantum Field Theories. In Sec.~\ref{sec:prac} we present some practical examples, and check with (2-dimensional) numerical integration of the RGE in the complex $Q^2$-plane that the algebraic procedure gives us the correct answer. In Sec.~\ref{sec:summ} we summarize our results.

\section{Implicit solution of the renormalization group equation}
\label{sec:impl}

The renormalization group equation (RGE) for the coupling parameter $F(z) \equiv a(Q^2) \equiv \alpha_s(Q^2)/\pi$, where $z \equiv \ln(Q^2/Q^2_{\rm in})$ is in general complex (and the initial scale is $Q^2_{\rm in} > 0$), can be written in the following way:
\be
\frac{d F(z)}{d z} = \beta(F(z)) \ ,
\label{RGE1}
\ee
where the beta function $\beta(F)$ characterizes a mass independent renormalization scheme in perturbative QCD (pQCD), i.e., it has a well defined expansion around $F=0$
\be
\beta(F)_{\rm exp} = - \beta_0 F^2 - \beta_1 F^3 - \beta_2 F^4 - \ldots \ .
\label{betexp}
\ee
Here, $\beta_0$ and $\beta_1$ are universal constants, $\beta_0=(11 - 2 n_f/3)/4$ and $\beta_1=(102-38 n_f/3)/16$, where $n_f$ is the number of active quark flavours. In the low-momentum regime ($|Q^2| \alt 10^1 \ {\rm GeV}^2$), this number is usually taken to be $n_f=3$, corresponding to the three lightest, almost massless, active quarks $u$, $d$ and $s$. The coefficients $\beta_j$ ($j \geq 2$) characterize the pQCD renormalization scheme \cite{Stevenson}.

As mentioned in the Introduction, there exist several theoretical arguments which suggest that the running coupling $F(z)   \equiv a(Q^2)$ is a finite function for all positive $Q^2$ and that it possibly acquires a finite positive value in the infared limit, $0 < a(0) \equiv a_0 < + \infty$. In this case, it turns out that $\beta(F)$ for positive couplings $F \equiv a$ has a root at $F=a_0$ [and double root at $F=0$ according to Eq.~(\ref{betexp})], and for $0 < F < a_0$ it has no roots. In view of this, we will consider the following class  of beta functions:
\be
\beta(F) \left( \equiv \beta(F)_{[M/N]} \right)
= - \beta_0 F^2 (1 - Y) \frac{T_M(Y)}{U_N(Y)} {\bigg \vert}_{Y \equiv F/a_0} \ ,
\label{beta1}
\ee
where $T_M(Y)$ and $U_N(Y)$ are polynomials of degree $M$ and $N$, respectively, both normalized in such a way that $T_M(0)=1=U_N(0)$. Specifically, we denote as $1/t_j$ the roots of $T_M(Y)$, and $1/u_j$ the roots of $U_N(Y)$
\bes
\label{roots}
\bea
T_M(Y) &=& (1 - t_1 Y) \cdots (1 - t_M Y) \ ,
\label{TM}
\\
U_N(Y) &=& (1 - u_1 Y) \cdots (1 - u_N Y) \ . 
\label{UN}
\eea
\ees
The parameters $t_j$ and $u_k$ are such that the polynomials $T_M(Y)$ and $U_N(Y)$ have real coefficients; this means that some of these parameters $t_j$ and $u_k$ can be real, and others complex conjugate pairs. 
We will restrict ourselves, for physical reasons, to such beta functions of the form (\ref{beta1}) in which those $t_j$ and $u_k$ which are real and positive are all below unity: $0 < t_j < 1$ and $0 < u_k < 1$. This means that:
\begin{itemize}
\item
$a=a_0$ is the smallest positive root of the beta function; 
\item
and that all those poles of the beta function which are positive are larger than $a_0$. 
\end{itemize}
 If the latter conditions were not fulfilled, the running coupling $a(Q^2)$ would obviously have (Landau) singularities on the positive $Q^2$-axis, contradicting the theoretical arguments mentioned in the Introduction. The former condition only means that we define $a_0$ as the smallest positive root of the beta function, and demand that at  least one such positive root exist. An important practical consequence of these restrictions will be highlighted in Sec.~\ref{sec:sing} (the first paragraph).

The first universal coefficient $\beta_0$ in the expansion of the beta function (\ref{betexp}) is reproduced automatically by our construction. The 
second universal coefficient $\beta_1$ in Eq.~(\ref{betexp}) imposes the following restriction on the polynomials $T_M(Y)$ and $U_N(Y)$:
\be
- \sum_{j=1}^M t_j + \sum_{k=1}^N u_k = 1 + \beta_1 a_0/\beta_0 \ ,
\label{beta1con}
\ee 
In addition, we will restrict the considered class of meromorphic beta functions to $M  +1 \geq N$. In such a case, it turns out that the RGE (\ref{RGE1}) can be integrated algebraically and leads to an implicit solution of the form $z = G(F)$ [for $F \equiv F(z)$]. Namely, the integration of the RGE (\ref{RGE1}) gives
\be
z = \int_{a_{\rm in}}^F \frac{ d F'}{\beta(F')} \ ,
\label{RGE2}
\ee
and if we introduce a new integration variable $t \equiv a_0/F$, this can be written as
\be
z = \frac{1}{\beta_0 a_0} \int_{a_0/a_{\rm in}}^{a_0/F} dt \frac{t \ U_N(1/t)}{(t-1) T_M(1/t)} \ ,
\label{RGE3}
\ee
where $a_{\rm in} = a(Q^2_{\rm in}) = F(z=0)$ has a real positive value, $0 < a_{\rm in} < a_0$.
When $M + 1 \geq N$, the integrand can be written as a sum of simple partial fractions $1/(t - t_j)$, where $t_0=1$ and $t_j$ ($j=1,\ldots,M$) are the roots of the ($M$-degree) polynomial $t^M T_M(1/t)$
\be
t^M T_M(1/t) \equiv t^M (1 - t_1/t) \cdots (1 - t_M/t) = (t - t_1) \cdots (t - t_M) \ .
\label{tMTM}
\ee
Namely, we have
\bes
\label{int}
\bea
\frac{t U_N(1/t)}{(t-1) T_M(1/t)} &=& \frac{t^{M+1} U_N(1/t)}{(t - t_0) (t - t_1) \cdots (t - t_M)}
\label{inta}
\\
& = & \left[ 1 + \sum_{j=0}^M B_j \frac{1}{(t - t_j)} \right] \ ,
\label{intb}
\eea
\ees
where the $M+1$ constants $B_j$ are 
\bea
B_j & = & \frac{t_j^{M+1} U_N(1/t_j)}{(t_j - t_0) \cdots (t_j - t_{j-1})(t_j - t_{j+1}) \cdots (t_j - t_M)} \ .
\label{Bj}
\eea
As a special case, we see that
\be
B_0=U_N(1)/T_M(1),
\label{B0}
\ee
which is a real number.
Using this, and the expression (\ref{beta1}), we also obtain the following relation:
\be
\beta^{'}(a)|_{a=a_0} = \beta_0 a_0 \frac{T_M(1)}{U_N(1)} = \frac{\beta_0 a_0}{B_0}.
\label{betapr}
\ee
Incidentally, in the limit of large $t$ the relations (\ref{int}) imply the following sum rule:
\be
\sum_{j=0}^M B_j = 1 - t_1^{(M)} + u_1^{(N)} = - \frac{\beta_1 a_0}{\beta_0},
\label{sumBj}
\ee
where the last equality is obtained by using the relation (\ref{beta1con}).
Using the form (\ref{intb}) for the integrand in Eq.~(\ref{RGE3}) leads us immediately to the implicit solution of the RGE
\be
z = \frac{1}{\beta_0 a_0} \left[ \left( \frac{a_0}{F(z)} - \frac{a_0}{a_{\rm in}} \right) + \sum_{j=0}^M B_j \ln \left(  \frac{a_0/F(z) - t_j}{a_0/a_{\rm in} - t_j} \right) \right].
\label{implsol}
\ee
Each logarithm has an ambiguity (winding number) because
\be
\ln A = \ln _{(\rm pb)} A +  i 2 \pi n_A = \ln |A| + i {\rm Arg}(A) + i 2 \pi n_A, \quad (n_A=0, \pm 1, \pm 2, \ldots),
\label{logWN}
\ee
where $\ln_{(\rm pb)}$ is the principal branch: $- \pi < {\rm Im} \ln_{(\rm pb)} A = {\rm Arg}(A) \leq + \pi$; further, $n_A$ is the winding number representing the ambiguity. When $A$ is positive, we consider that $\ln A$ is automatically the principal branch.
This would then suggest that the right-hand side of Eq.~(\ref{implsol}) has $M+1$ independent winding numbers $n_j$, correspondig to each logarithm there. The physically acceptable winding numbers of the logarithms on the right-hand side of Eq.~(\ref{implsol}) are such that they give for the expression (\ref{implsol}) a number $z = \ln(Q^2/Q^2_{\rm in})$ corresponding to the squared impulse $Q^2$ on the first Riemann sheet, i.e., $| {\rm Im} z| \leq \pi$ (cf.~also the discussion in the beginning of Sec.~\ref{subs:Lpole}). 

However, in general some of the roots $t_j$ of the polynomial $T_M(Y)$, Eq.~(\ref{TM}), are not real, but form complex conjugate pairs. For example, if the first complex conjugate pair is $(t_1, t_2=t_1^{\ast})$, then it is straightforward to check that the corresponding coefficients $B_1$ and $B_2$ are mutually complex conjugate, and the corresponding two terms in the sum (\ref{intb}) are
\bea
B_1 \frac{1}{(t-t_1)} + B_2 \frac{1}{(t - t_2)} & = &
2 \; \frac{ t {\rm Re} (B_1) - {\rm Re}(t_1^{\ast} B_1) }{t^2 - 2 t {\rm Re}(t_1) + |t_1|^2}
\label{Btpair}
\eea
and the coresponding contribution to the integral (\ref{RGE3}) is
\bea
\lefteqn{
  \frac{1}{\beta_0 a_0} \int_{a_0/a_{\rm in}}^{a_0/F} dt \; 2 \; \frac{ t {\rm Re} (B_1) - {\rm Re}(t_1^{\ast} B_1) }{t^2 - 2 t {\rm Re}(t_1) + |t_1|^2} =
}
\nonumber \\ &&
\frac{1}{\beta_0 a_0} {\Bigg \{} 
\frac{2
\left( ({\rm Re} B_1) ({\rm Re} t_1) - \rm{Re} (t_1^{\ast} B_1) \right)}{| {\rm Im}  t_1 |}
\left[ {\rm ArcTan} \left( \frac{ a_0/F(z) - {\rm Re} t_1}{| {\rm Im}  t_1 |} \right) - {\rm ArcTan} \left( \frac{ a_0/a_{\rm in} - {\rm Re} t_1}{| {\rm Im}  t_1 |} \right) \right]
\nonumber \\
&& + ({\rm Re} B_1) \left[
\ln \left( \left(\frac{a_0}{F(z)}\right)^2 - 2 ({\rm Re} t_1) \frac{a_0}{F(z)} + |t_1|^2 \right) - 
\ln \left( \left(\frac{a_0}{a_{\rm in}}\right)^2 - 2 ({\rm Re} t_1) \frac{a_0}{a_{\rm in}} + |t_1|^2 \right) \right] 
{\Bigg \}}.
\label{IntBtpair}
\eea
Therefore, the expression on the right-hand side of Eq.~(\ref{implsol}) can be rewritten more explicitly, for the case when $t_j$ ($j=1,\ldots, 2 P$) are $P$ complex conjugate pairs and $t_j$ ($j=2 P +1, \ldots, M$) are real
\bea
\lefteqn{
z = \frac{1}{\beta_0 a_0} {\Bigg \{} \left( \frac{a_0}{F(z)} - \frac{a_0}{a_{\rm in}} \right) +  B_0 \ln \left(  \frac{a_0/F(z) - 1}{a_0/a_{\rm in} - 1} \right)
+ \sum_{j=2 P + 1}^M B_j \ln \left(  \frac{a_0/F(z) - t_j}{a_0/a_{\rm in} - t_j} \right)}  
\nonumber\\ &&
+ \sum_{k=0}^{P-1} \frac{2
\left( ({\rm Re}B_{2 k+1}) ({\rm Re} t_{2 k +1}) - {\rm Re} (t_{2 k +1}^{\ast} B_{2 k +1}) \right)} {| {\rm Im}  t_{2 k +1} |}
\left[ {\rm ArcTan} \left( \frac{ a_0/F(z) - {\rm Re} t_{2 k +1}}{| {\rm Im}  t_{2k +1} |} \right) -
  {\rm ArcTan} \left( \frac{ a_0/a_{\rm in} - {\rm Re} t_{2 k +1}}{| {\rm Im}  t_{2k +1} |} \right) \right]
\nonumber\\
&& + \sum_{k=0}^{P-1} ({\rm Re} B_{2 k +1}) \left[
\ln  \left( \left(\frac{a_0}{F(z)}\right)^2 - 2 ({\rm Re} t_{2 k +1}) \frac{a_0}{F(z)} + |t_{2 k +1}|^2 \right) - 
\ln  \left( \left(\frac{a_0}{a_{\rm in}}\right)^2 - 2 ({\rm Re} t_{2 k +1}) \frac{a_0}{a_{\rm in}} + |t_{2 k +1}|^2 \right)
\right]
{\Bigg \}}.
\label{implsolg}
\eea
We note that among the $P$ $z$-dependent ${\rm ArcTan}$ terms, which are in general complex, each has one winding number because for $A = |A| \exp(i \theta)$ ($|\theta| \leq \pi$)\footnote{The $z$-independent ${\rm ArcTan}$ terms in Eq.~(\ref{implsolg}) are real (because $a_{\rm in}$ is real, $0< a_{\rm in} < a_0$).}
\be
{\rm ArcTan} (A) = {\rm ArcTan}_{(\rm pb)} (A) + \pi n_A, \quad (n_A=0, \pm 1, \pm 2, \ldots),
\label{ArcTanWN}
\ee
where we regard as the principal branch ${\rm ArcTan}_{(\rm pb)} (A)$ the one which fulfills the inequality $- \pi/2 < {\rm Re} {\rm ArcTan}_{(\rm pb)} (A) \leq + \pi/2$. When $A$ is real, we consider that ${\rm ArcTan}$ is automatically the principal branch.

Further, each of the $M-P+1$ $z$-dependent logarithms appearing in Eq.~(\ref{implsolg}) has a winding number according to the relation (\ref{logWN}).
This means that we have in general in total $M+1$ winding numbers. This realization will play a role in the next Section \ref{sec:sing}. 

We recall that the considered $\beta(a)$ functions are such that $a(Q^2)$ is a holomorphic function in and around any positive point $Q^2>0$. However, at $Q^2=0$, where $a=a_0 < \infty$, the function $a(Q^2)$ could be nonholomorphic (nonanalytic), i.e., certain (high enough) derivative $(d/d Q^2)^n a(Q^2)$ at $Q^2=0$ could be infinite. In our considered cases we have for the Taylor expansion around $Q^2=0$
\be
a(Q^2) = a_0 + C_0 \left( \frac{Q^2}{\Lambda^2} \right)^{\kappa} + C_1  \left( \frac{Q^2}{\Lambda^2} \right)^{2 \kappa} + \ldots
\label{aQ2exp}
\ee
This implies
\bes
\bea
\beta(a(Q^2)) & = & \kappa (a(Q^2) - a_0) + {\cal O}\left((a(Q^2)-a_0)^2 \right),
\label{betaexp}
\\
\Rightarrow \; \beta^{'}(a)|_{a=a_0} &=& \kappa .
\label{kappa0}
\eea
\ees
The use of relation (\ref{betapr}) then gives the power index $\kappa$ in terms of the parameters contained in the considered beta function Eq.~(\ref{beta1})
\be
\kappa= \beta_0 a_0 \frac{T_M(1)}{U_N(1)} = \frac{\beta_0 a_0}{B_0}.
\label{kappa}
\ee
In general, $\kappa$ is noninteger, and consequently the coupling is in general not analytic at $Q^2=0$ ($z=-\infty$).

We wish to point out that the class of the $\beta$-functions considered here, Eq.~(\ref{beta1}), in addition to having a Pad\'e form $P[M+3/N](a)$, have specific restrictions which result in a finite positive and monotonically decreasing running coupling $a(Q^2) < a_0$ on the entire nonnegative $Q^2$-axis $Q^2 \geq 0$ [with $a(Q^2) \to a_0$ when $Q^2 \to 0$]. This is reflected in the formal requirement that those of the parameters $t_j$ and $u_k$ of Eqs.~(\ref{roots}) which are not complex and are positive must fulfill the restrictions $0 < u_k < 1$ and $0 < t_j < 1$.

On the other hand, there are specific classes of Pad\'e-type QCD $\beta$-functions which do not fulfill the above restrictions [i.e., they do not give finite $a(Q^2)$ on the entire positive $Q^2$-axis],\footnote{
Stated otherwise, there are Landau singularities on the positive $Q^2$-axis in such cases.} but give explicit solutions $a(Q^2)$ of the RGE where $a(Q^2)$ involves the Lambert function $W$. Specifically, when $\beta(a)$ is of a Pad\'e-form $P[2/1](a)$ such that it reproduces the correct $\beta_j$-coefficients up to two-loop ($\beta_0, \beta_1$) \cite{Gardi,Magradze,Jamin}\footnote{In the context of the ${\cal N}=1$ supersymmetric Yang-Mills theory, cf.~\cite{NSVZ,Jones,Kataev}.}; when $\beta(a)$ is of a Pad\'e-form $P[3/1](a)$ such that it reproduces the chosen $\beta_j$-coefficients up to three-loop ($\beta_j$, $j=0,1,2$) \cite{Gardi}; when $\beta(a)$ is of a Pad\'e-form $P[4/4](a)$ reproducing the chosen $\beta_j$-coefficients up to four-loop ($\beta_j$, $j=0,1,2, 3$) \cite{GCIK};\footnote{For a practical application, in a specific (MiniMOM) scheme, cf.~\cite{3dAQCD}.} and even up to five-loop \cite{GCIK} [in that case $\beta(a)$ is of a Pad\'e-form $P[5/6](a)$].

In the considered class of $\beta$-functions (\ref{beta1})-(\ref{roots}), with the mentioned restrictions $t_j < 1$ and $u_k <1$ when $t_j$ or $u_k$ are real, the following question may arise: when expanding the $\beta$-function in powers of $F$, Eq.~(\ref{betexp}), which values of the renormalization scheme parameters $c_n \equiv \beta_n/\beta_0$ ($n \geq 2$) can be generated? Direct expansion gives the relations
\be
c_n^{(r)} \equiv a_0^n c_n = \sum_{s=0}^n (-1)^{n-s} \sum t_{j_1} \cdots t_{j_{n-s}} \sum u_{k_1} \cdots u_{k_s},
\label{cn}
\ee
where the sum is over $0 \leq j_1 < \ldots < j_{n-s} \leq M$ (taking $t_0=1$) and
$1 \leq k_1 \leq \ldots \leq k_s \leq N$. For $n=1$ this relation reduces to the condition (\ref{beta1con}), where $c_1 \equiv \beta_1/\beta_0$ is a universal coefficient ($c_1=51/22$ when $n_f=0$; $c_1=16/9$ when $n_f=3$). In a considered $\beta$-function form (\ref{beta1}), for chosen $M$ and $N$ and a chosen value of $a_0 \equiv a(0) > 0$, the relation (\ref{beta1con}) relates the $(M+N)$ parameters $t_j$ ($1 \leq j \leq M$) and $u_k$ ($1 \leq k \leq N$), and consequently we have $(M+N-1)$ degrees of freedom (d.o.f.). These $(M+N-1)$ d.o.f. then give us the first $(M+N-1)$ independent scheme parameters $c_2, \ldots, c_{M+N}$. It turns out that in general the values of the latter scheme parameters cover the entire real axis (i.e., all the values) once the $(M+N-1)$ independent coefficients $t_j$ and $u_k$ are varied across all the allowed range; the only exception may be the last scheme coefficient $c_{M+N}$ which may vary only over a part of the real axis.

For example, if $M+N=2$, only the first scheme parameter $c_2 \equiv \beta_2/\beta_0$ is independent. More specifically, there are three cases 
\bes
\label{restr}
\bea
M=N=1 & \Rightarrow & c_2 < \frac{3 c_1}{a_0} ,
\label{MN11}
\\
M=2 \; \& \; N=0 & \Rightarrow & -\frac{1}{a_0} \left( \frac{3}{a_0} + 2 c_1 \right) < c_2.
\label{MN20}
\\
M=0 \; \& \; N=2 & \Rightarrow & c_2 < {\rm Max} \left[ \frac{1}{4} \left( \frac{1}{a_0} + c_1 \right) \left( - \frac{1}{a_0} + 3 c_1 \right), \; c_1^2 \right]
\label{MN02}
\eea
\ees
When $M=N=1$, the coefficients $t_1$ and $u_1$ must be real and are thus both below unity ($t_1, u_1 <1$), which gives us the restriction (\ref{MN11}). When $M=2$ and $N=0$, and $t_1$ and $t_2$ are mutually complex conjugate, the restriction on $c_2$ is $(1/4) (1/a_0 +c_1) (-3/a_0 + c_1) \leq c_2$; and when $t_1$ and $t_2$ are real (and thus below unity), the restriction is $-(1/a_0) (+3/a_0 + 2 c_1) < c_2 < (1/4) (1/a_0 +c_1) (-3/a_0 + c_1)$; combining these two restrictions gives us the restriction (\ref{MN20}). When $M=0$ and $N=2$, a similar analysis leads to the restriction (\ref{MN02}).

The infrared limit $a(0) \equiv a_0$ ($>0$) can be, in principle any number. Nonetheless, the QCD phenomenology requires in practice that $a_0 > (a_0)_{\rm min}$. In such case, in all the restrictions (\ref{restr}) we must replace $a_0$ by $(a_0)_{\rm min}$.

We can see from the restrictions (\ref{restr}) that in the case $M+N=2$ the first (two-loop) scheme parameter $c_2$ covers all possible (real) values once we allow, for example, in addition to the form $M=N=1$ also the form $M=2$ and $N=0$.

Alternatively, if enlarge the $M=N=1$ form to the form $M=2$ and $N=1$, we can also see that this generates all possible values of $c_2$ (and a restricted range of values of $c_3$).

In general, for the first $(n-1)$ scheme parameters $c_2, \ldots, c_n$ (where $n \geq 2$ is fixed), all their values can be generated if we consider a sufficiently large set of $\beta$-functions of the type (\ref{beta1}), i.e., with various choices for the values of the indices $M$ and $N$ and with full variation of the parameters $t_j$ and $u_k$ under the mentioned restrictions.

\section{Singularities (Landau) outside the real $Q^2$-axis}
\label{sec:sing}

We note that, by restrictions on the beta function mentioned in the previous Section, the running coupling $a(Q^2) \equiv F(z)$ has no singularities on the real positive $Q^2$-axis, i.e., there are no positive-$Q^2$ Landau singularities. This is so because $a(Q^2)$ is constrained there to run between  the value $a(0) = a_0$ ($> 0$) and $a(+\infty) = 0$, the latter equality being valid by the asymptotic freedom of QCD reflected in the form (\ref{betexp}) of beta function when $F \to 0$. Namely, by our restrictions on the class of considered $\beta(a)$ functions, when $a(Q^2)$ is RGE-running with increasing positive $Q^2$, beta function $\beta(a(Q^2))$ will be negative finite all the time, since no new roots or poles of the beta function are encountered. Therefore, by the mentioned restrictions on the roots and poles of the beta function (\ref{beta1}) we ensured in advance that the positive-$Q^2$ Landau singularities (poles and/or cuts) do not exist.\footnote{See also Fig.~\ref{Figa1all} at the end of Sec.~\ref{sec:prac} for three representative cases of the running of $a(Q^2)$ for $Q^2>0$.} 

\subsection{Landau poles}
\label{subs:Lpole}

\begin{figure}[htb] 
\centering\includegraphics[width=140mm]{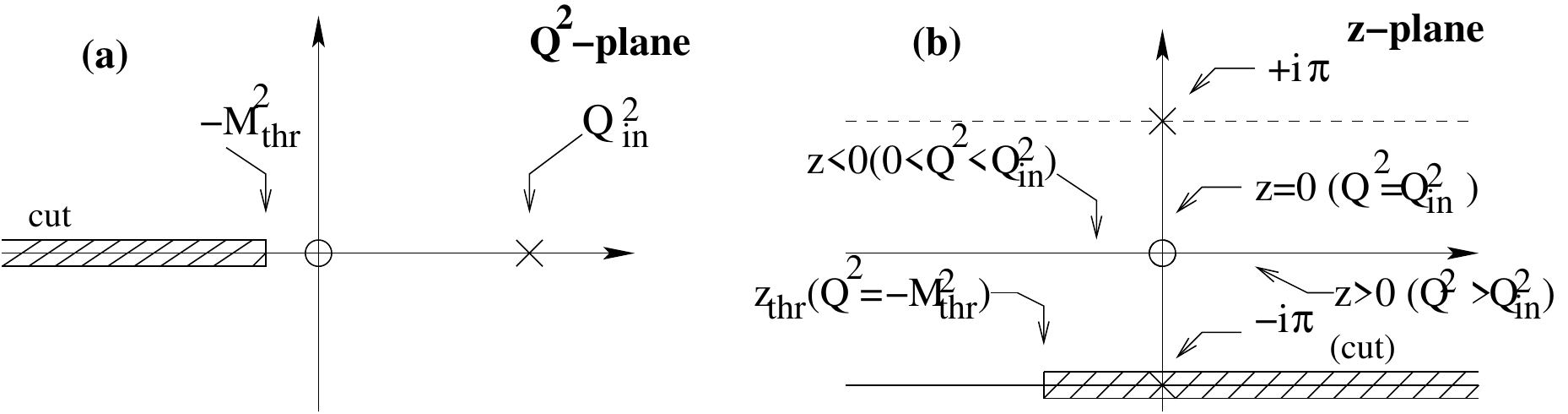}
\vspace{-0.4cm}
 \caption{\footnotesize  (a) Complex $Q^2$-plane; (b) complex
$z$-plane where $z=\ln(Q^2/Q_{\rm in}^2)$; the physical
stripe is $-\pi \leq {\rm Im}z < + \pi$.}
\label{Qzplane}
 \end{figure}

We will now construct an algebraic algorithm which allows us to verify whether in the (first Riemann sheet of the) complex $Q^2$-plane the solution (\ref{implsol}) has poles outside the real $Q^2$-axis (Landau poles). We assume that only the first sheet of the complex $Q^2$-plane has physical meaning,\footnote{
\label{1stRSfn}This assumption is related with the usual dispersive integral representation of the coupling $a(Q^2)=F(z)$, which is applicable in the first Riemann sheet.}
i.e., $Q^2 = |Q^2| \exp(i \phi)$ where $- \pi \leq \phi < + \pi$.  This corresponds for the $z \equiv \ln(Q^2/Q^2_{\rm in})$ variable to be a band in the complex $z$-plane with $- \pi \leq {\rm Im} z < + \pi$, cf.~Figs.~\ref{Qzplane} (a), (b).
As argued in the Introduction, if $a(Q^2)$ is to reflect the holomorphic properties of spacelike Green functions and observables, such as current correlators  or structure functions, then $a(Q^2)$ can have singularities (cut) only along the negative $Q^2$-axis: $-\infty < Q^2 \leq - M^2_{\rm thr}$, where the threshold mass $M^2_{\rm thr}$ is either positive ($\sim 0.1 \ {\rm GeV}^2$) or zero. This cut corresponds in the $z$-stripe to the cut along the ${\rm Im} z = - \pi$ border line. 

As explained, the possible Landau singularities in the considered pQCD renormalization schemes are within the $Q^2$-complex plane outside the real $Q^2$-axis. In the $z$-plane this corresponds to the possible Landau singularities within the interior of the $z$-stripe, $-\pi < z < +\pi$, but not along the real axis, $z \not\in \mathbb{R}$.

Landau pole $z_{\ast} = x_{\ast} + i y_{\ast}$ [$\Leftrightarrow Q^2_{\ast} = Q^2_{\rm in} \exp(x_{\ast}) \exp(i y_{\ast})$] is usually a branching point of a cut singularity of $F(z)$, such that $F(z_{\ast})=\infty$, and it is situated on the first Riemann sheet outside the timelike semiaxis ($|{\rm Im} z_{\ast}| < \pi$). Let us denote $a_{\ast} \equiv F(x_{\ast})$  ($0 < a_{\ast} < a_0$). We then apply the implicit solution Eq.~(\ref{implsolg}) to the points $z_1 = x_{\ast}$ and $z_2=x_{\ast} + i y_{\ast}$, and subtract the two equations; this then gives us the equation
\be
y_{\ast} =  {\cal G}_{\vec n}(a_{\ast})
\label{Lp}
\ee
where
\bea
{\cal G}_{\vec n}(a_{\ast}) &\equiv& \frac{(-i)}{\beta_0 a_0} {\Bigg \{} - \frac{a_0}{a_{\ast}}  +  B_0 \left[ i \pi - \ln \left(\frac{a_0}{a_{\ast}} - 1 \right) \right]
+ \sum_{j=2 P + 1}^M B_j \left[ \ln_{\rm (pb)} (- t_j) - \ln \left( \frac{a_0}{a_{\ast}}  - t_j \right) \right]
\nonumber\\ &&
+ \sum_{k=0}^{P-1} \frac{2
\left( ({\rm Re}B_{2 k+1}) ({\rm Re} t_{2 k +1}) - {\rm Re} (t_{2 k +1}^{\ast} B_{2 k +1}) \right)} {| {\rm Im}  t_{2 k +1} |}
\left[ {\rm ArcTan} \left( \frac{ - {\rm Re} t_{2 k +1}}{| {\rm Im}  t_{2k +1} |} \right) -
  {\rm ArcTan} \left( \frac{ a_0/a_{\ast} - {\rm Re} t_{2 k +1}}{| {\rm Im}  t_{2k +1} |} \right) \right]
\nonumber\\
&& + \sum_{k=0}^{P-1} ({\rm Re} B_{2 k +1}) \left[
\ln |t_{2 k +1}|^2  - 
\ln  \left( \left(\frac{a_0}{a_{\ast}}\right)^2 - 2 ({\rm Re} t_{2 k +1}) \frac{a_0}{a_{\ast}} + |t_{2 k +1}|^2 \right)
\right] {\Bigg \}}
\nonumber\\
&& 
+ \frac{2 \pi}{\beta_0 a_0} \left( B_0 n_0 + \sum_{j=2 P +1}^M B_j n_j \right)
\nonumber\\
&& 
+ \frac{2 \pi}{\beta_0 a_0} \sum_{k=0}^{P-1} ({\rm Re} B_{2 k +1}) N_k
+ i \frac{2 \pi}{\beta_0 a_0} \sum_{k=0}^{P-1} \frac{ \left( {\rm Re}(t_{2 k +1}^{\ast} B_{2 k +1}) - ({\rm Re} B_{2 k+1})({\rm Re} t_{2 k+1}) \right)}{| {\rm Im}  t_{2 k+ 1} |} {\cal N}_k.
\label{Ga}
\eea
Here, we accounted for the nonuniqueness of the ($z$-dependent) logarithms Eq.~(\ref{logWN}) and ArcTan Eq.~(\ref{ArcTanWN})
\bes
\label{limWN}
\bea
\lim_{y \to y_{\ast}} \ln \left( \frac{a_0}{F(x_{\ast} + i y)} - t_j \right) & = &
\ln _{(\rm pb)}(- t_j) + i 2 \pi n_j \;  (j=0; 2 P+1, \ldots, M),
\label{limLnWNa}
\\
\lim_{y \to y_{\ast}} \ln \left[ \left(\frac{a_0}{F(x_{\ast} + i y)} \right)^2 - 2 ({\rm Re} t_{2 k+1}) \frac{a_0}{F(x_{\ast} + i y)}  + |t_{2 k+1}|^2 \right] & = &
\ln |t_{2 k+1}|^2 + i 2 \pi N_k \quad (k = 0, \ldots, P-1),
\label{limLnWNb}
\\
\lim_{y \to y_{\ast}} {\rm ArcTan} \left( \frac{ \frac{a_0}{F(x_{\ast} + i y)} - {\rm Re} t_{2 k + 1} }{|{\rm Im} t_{2 k+1} |} \right)  & = & {\rm ArcTan} \left( \frac{ - {\rm Re} t_{2 k + 1} }{|{\rm Im} t_{2 k+1} |} \right) + \pi {\cal N}_k,
\label{limArcTanWN}
\eea
\ees
where $(k=0,\ldots,P-1)$, and we denoted the $M+1$ winding numbers
\be
{\vec n} \equiv \{ n_0,n_{2 P+1},\ldots, n_M; N_0,\ldots, N_{P-1}; {\cal N}_0, \ldots, {\cal N}_{P-1} \},
\label{ndef}
\ee
where $n_j, N_k, {\cal N}_k = 0, \pm 1, \pm 2, \ldots$. We note that the terms $\ln _{(\rm pb)}(- t_j)$ may have $t_j$ either negative or positive, and therefore
\be
\ln _{(\rm pb)}(- t_j) = \ln |t_j| + \Theta(t_j) i \pi.
\label{lnP}
\ee
As a special case, we used in Eq.~(\ref{Ga}):  $\ln(-t_0)=\ln(-1) = i \pi$.
We note that, since the real roots $t_j$ fulfill the inequality  $t_j \leq 1$ [our initial physical restrictions on $\beta$-function, cf.~comments after Eqs.~(\ref{roots})], the logarithm $\ln (a_0/a_{\ast} - t_j)$ in Eq.~(\ref{Ga}) is a real number because it has positive argument. For the same reason, also the $P$ logarithms of the trinomials in $(a_0/a_{\ast})$ in Eq.~(\ref{Ga}) are real.
We point out that that winding numbers ${\vec n}$ appear when integrating the RGE (\ref{RGE1}) from $z_1=x_{\ast}$ to $z_2=x_{\ast} +  i y{\ast}$. 

Equation (\ref{Lp}) for the poles represents two equations, one for the imaginary and one for the real parts
\be
{\rm Im} {\cal G}_{\vec n}(a_{\ast}) = 0, \quad
{\rm Re} {\cal G}_{\vec n}(a_{\ast}) = y_{\ast},    
\label{LpReIm}
\ee
where
\bes
\label{ImReG}
\bea
{\rm Im} {\cal G}_{\vec n}(a_{\ast}) & \equiv &
 \frac{1}{\beta_0 a_0} {\Bigg \{}  \frac{a_0}{a_{\ast}}  +  B_0 \ln \left(\frac{a_0}{a_{\ast}} - 1 \right) 
+ \sum_{j=2 P + 1}^M B_j \left[ -\ln_{\rm (pb)} |t_j| + \ln \left( \frac{a_0}{a_{\ast}}  - t_j \right) \right]
\nonumber\\ &&
+ \sum_{k=0}^{P-1} \frac{2
\left( {\rm Re} (t_{2 k +1}^{\ast} B_{2 k +1}) - ({\rm Re}B_{2 k+1}) ({\rm Re} t_{2 k +1}) \right)} {| {\rm Im}  t_{2 k +1} |}
\left[- {\rm ArcTan} \left( \frac{ - {\rm Re} t_{2 k +1}}{| {\rm Im}  t_{2k +1} |} \right)  +
  {\rm ArcTan} \left( \frac{ a_0/a_{\ast} - {\rm Re} t_{2 k +1}}{| {\rm Im}  t_{2k +1} |} \right) \right]
\nonumber\\
&& + \sum_{k=0}^{P-1} ({\rm Re} B_{2 k +1}) \left[
- \ln |t_{2 k +1}|^2  +
\ln  \left( \left(\frac{a_0}{a_{\ast}}\right)^2 - 2 ({\rm Re} t_{2 k +1}) \frac{a_0}{a_{\ast}} + |t_{2 k +1}|^2 \right)
\right] {\Bigg \}}
\nonumber\\ &&
+ \frac{2 \pi}{\beta_0 a_0} \sum_{k=0}^{P-1} \frac{
  \left( {\rm Re} (t_{2 k +1}^{\ast} B_{2 k +1}) - ({\rm Re}B_{2 k+1}) ({\rm Re} t_{2 k +1}) \right)}
  {| {\rm Im}  t_{2 k +1} |} {\cal N}_k,
\label{ImG}
\\
{\rm Re} {\cal G}_{\vec n}(a_{\ast}) & \equiv &
\frac{2 \pi}{\beta_0 a_0} \left\{ B_0 \left( n_0 + \frac{1}{2} \right) + \sum_{j=2 P +1}^M B_j \left( n_j + \frac{1}{2} \Theta(t_j) \right) + \sum_{k=0}^{P-1} ({\rm Re} B_{2 k +1}) N_k \right\}.
\label{ReG}
\eea
\ees
We note that ${\rm Im} {\cal G}_{\vec n}(a_{\ast})$ depends only on the winding numbers ${\cal N}_k$ ($k=0,\ldots, P-1$) coming from ${\rm ArcTan}$; and ${\rm Re} {\cal G}_{\vec n}(a_{\ast})$ depends on the winding numbers $n_j$ ($j=0,2 P+1, 2 P+2, \ldots, M$) and $N_k$ ($=0,\ldots, P-1$), both coming from logarithms.

The necessary conditions for the existence of a Landau pole are
\begin{enumerate}
\item
${\rm Im} {\cal G}_{\vec n}(a_{\ast})=0$ for a chosen set ${\cal N}_k$ ($k=0,\ldots, P-1$), and the value $a_{\ast}$ lies between $0$ and $a_0$ ($0 < a_{\ast} < a_0$);
\item
and simultaneously, ${\rm Re} {\cal G}_{\vec n}(a_{\ast})$ ($=y_{\ast}$) is within the interior of the first Riemann sheet, i.e., inside the first stripe of $z$, $|{\rm Re} {\cal G}_{\vec n}(a_{\ast})| < \pi$, for certain choices of $n_j$ ($j=0,2 P+1, 2 P+2, \ldots, M$) and $N_k$ ($=0,\ldots, P-1$).
\end{enumerate}  

If, for example, all $B_j$ coefficients are real, then ${\rm Im} \; {\cal G}_{\vec n}(a) = {\rm Im} \; {\cal G}_{\vec 0}(a)$; if in such a case ${\rm Im} \; {\cal G}_{\vec 0}(a_{\ast})$ has no zero in the positive interval $0 < a_{\ast} < a_0$, then one necessary condition for the existence of Landau poles is not fulfilled, i.e., there are no Landau poles.

\subsection{Landau branching points}
\label{subs:Lbranch}

In the previous Subsection we presented an algorithm which allows us to find, inside the complex $z$-stripe, the (Landau) poles where the coupling is infinite $F(z_{\ast})= \infty$. However, the complex function $F(z)$ can have also another type of Landau singularities, namely a cut with a finite-valued branching point $z_{\ast}$.

One illustrative mathematical example is $F(z) = (z - z_{\ast})^{1/2}$, where $z_{\ast} = x_{\ast} + i y_{\ast}$ is such a branching point, $F(z_{\ast}) = 0$ and $F'(z) = \infty$. The cut in this case is usually defined along the semiaxis to the left of $z_{\ast}$: $x + i y_{\ast}$ ($x \leq x_{\ast}$).

However, we may worry at first that other, even more ``finite,'' type of Landau branching points $z_{\ast} \not\in \mathbb{R}$ may appear, such as $F(z) = (z - z_{\ast})^{3/2}$, for which $F'(z_{\ast}) < \infty$ and $F''(z_{\ast}) = \infty$. We show that this does not occur for the considered class of meromorphic beta functions (\ref{beta1})-(\ref{roots}). Namely,
\bea
F''(z) & = & \frac{d}{d z} \beta(F(z)) = \beta(F(z)) \frac{\partial}{\partial F} \beta(F(z)) \ .
\label{Fpp}
\eea
The poles of the right-hand side are at the same values $F=a_0/u_s$ ($s=1,\ldots,N$) as in the beta function $\beta(F)$ itself, cf.~Eqs.~(\ref{beta1})-(\ref{roots}). This means that, if $F''(z_{\ast}) = \infty$, then $F'(z_{\ast})=\infty$. We can continue this argumentation, by applying further derivatives $(d/dz)^n$ to Eq.~(\ref{Fpp}). E.g., if $F^{(3)}(z_{\ast})=\infty$, then $F'(z_{\ast})=\infty$.

Therefore, the only relevant situation of finite-valued Landau branching points $z_{\ast}$ for the considered beta functions is: $F'(z_{\ast})=\infty$ and $F(z_{\ast}) < \infty$. Since $F'(z_{\ast}) = \beta(F(z_{\ast}))$, such a branching point is one of the poles of the beta function, $z_{\ast}^{(s)} = x_{\ast}^{(s)} + i y_{\ast}^{(s)}$ such that $F(z_{\ast}^{(s)})= a_0/u_s$ ($s=1,\ldots,N$), cf.~Eqs.~(\ref{beta1})-(\ref{roots}). This means, in analogy with Eqs.~(\ref{Lp})-(\ref{Ga}) and using the notations (\ref{ndef}), that we have the relation
\be
y_{\ast}^{(s)}  = {\cal K}_{\vec n}(a_{\ast}^{(s)}; u_s) \ ,
\label{Lbp}
\ee
where $a_{\ast}^{(s)} \equiv F(x_{\ast}^{(s)})$  ($0 < a_{\ast}^{(s)} < a_0$), and
\bea
\lefteqn{
{\cal K}_{\vec n}(a_{\ast}^{(s)}; u_s) \equiv   \frac{(-i)}{\beta_0 a_0} {\Bigg \{}
\left( u_s - \frac{a_0}{a_{\ast}^{(s)}} \right) +  B_0 \left[  \ln_{(\rm pb)} (u_s - 1) - \ln \left( \frac{a_0}{a_{\ast}^{(s)}} - 1 \right) \right]
}
\nonumber\\
&&
+ \sum_{j=2 P + 1}^M B_j \left[  \ln_{(\rm pb)} (u_s - t_j) - \ln \left( \frac{a_0}{a_{\ast}^{(s)}} - t_j \right) \right]
\nonumber\\ &&
+ \sum_{k=0}^{P-1} \frac{2
\left( ({\rm Re}B_{2 k+1}) ({\rm Re} t_{2 k +1}) - {\rm Re} (t_{2 k +1}^{\ast} B_{2 k +1}) \right)} {| {\rm Im}  t_{2 k +1} |}
\left[ {\rm ArcTan} \left( \frac{ u_s - {\rm Re} t_{2 k +1}}{| {\rm Im}  t_{2k +1} |} \right) -
  {\rm ArcTan} \left( \frac{ a_0/a_{\ast}^{(s)} - {\rm Re} t_{2 k +1}}{| {\rm Im}  t_{2k +1} |} \right) \right]
\nonumber\\
&& + \sum_{k=0}^{P-1} ({\rm Re} B_{2 k +1}) \left[
\ln \left( u_s^2 - 2 ({\rm Re} t_{2 k +1}) u_s + |t_{2 k +1}|^2  \right) - 
\ln  \left( \left(\frac{a_0}{a_{\ast}^{(s)}}\right)^2 - 2 ({\rm Re} t_{2 k +1}) \frac{a_0}{a_{\ast}^{(s)}} + |t_{2 k +1}|^2 \right)
\right] {\Bigg \}}
\nonumber\\ &&
+ \frac{2 \pi}{\beta_0 a_0} \left( B_0 n_0^{(s)} + \sum_{j=2 P +1}^M B_j n_j^{(s)} \right)
\nonumber\\
&& 
+ \frac{2 \pi}{\beta_0 a_0} \sum_{k=0}^{P-1} ({\rm Re} B_{2 k +1}) N_k^{(s)}
+ i \frac{2 \pi}{\beta_0 a_0} \sum_{k=0}^{P-1} \frac{ \left( {\rm Re}(t_{2 k +1}^{\ast} B_{2 k +1}) - ({\rm Re} B_{2 k+1})({\rm Re} t_{2 k+1}) \right)}{| {\rm Im}  t_{2 k+ 1} |} {\cal N}_k^{(s)}.
\label{Ka}
\eea
The winding numbers are generated in a limiting process analogous to that in Eqs.~(\ref{limWN})
\bes
\label{limWNp}
\bea
\lefteqn{
\lim_{y \to y_{\ast}^{(s)}} \ln \left( \frac{a_0}{F(x_{\ast}^{(s)} + i y)} - t_j \right) = 
\ln (u_s - t_j)
}
\nonumber\\ 
&&
=\ln _{(\rm pb)}(u_s - t_j) + i 2 \pi n_j^{(s)} \quad (j=0; 2 P+1, 2 P + 2, \ldots, M),
\label{limLnWNap}
\\
\lefteqn{\lim_{y \to y_{\ast}^{(s)}} \ln \left[ \left(\frac{a_0}{F(x_{\ast}^{(s)} + i y)} \right)^2 - 2 ({\rm Re} t_{2 k+1}) \frac{a_0}{F(x_{\ast}^{(s)} + i y)}  + |t_{2 k+1}|^2 \right] =}
\nonumber\\ &&
\ln\left(u_s^2 - 2 ({\rm Re} t_{2 k+1}) u_s + |t_{2 k+1}|^2 \right) + i 2 \pi N_k^{(s)} \quad (k = 0, \ldots, P-1),
\label{limLnWNbp}
\\
\lefteqn{\lim_{y \to y_{\ast}^{(s)}} {\rm ArcTan} \left( \frac{ \frac{a_0}{F(x_{\ast}^{(s)} + i y)} - {\rm Re} t_{2 k + 1} }{|{\rm Im} t_{2 k+1} |} \right)  =  {\rm ArcTan} \left( \frac{ u_s - {\rm Re} t_{2 k + 1} }{|{\rm Im} t_{2 k+1} |} \right)
  }
\nonumber\\ &&
={\rm ArcTan} _{(\rm pb)} \left( \frac{ u_s - {\rm Re} t_{2 k + 1} }{|{\rm Im} t_{2 k+1} |} \right) + \pi {\cal N}_k^{(s)} \quad (k = 0, \ldots, P-1).
\label{limArcTanWNp}
\eea
\ees
As in Eqs.~(\ref{limWN})-(\ref{ndef}), the $M+1$ winding numbers appear
\be
{\vec n} \equiv \{ n_0^{(s)},n^{(s)}_{2 P+1},\ldots, n^{(s)}_M; N^{(s)}_0,\ldots, N^{(s)}_{P-1}; {\cal N}^{(s)}_0, \ldots, {\cal N}^{(s)}_{P-1} \},
\label{ndefp}
\ee
when integrating the RGE (\ref{RGE1}) in the $z$-plane  along the vertical direction, from $z_1=x_{\ast}^{(s)}$ toward $z_2=x_{\ast}^{(s)} + i y_{\ast}^{(s)}$.

We note that one of the physically motivated restrictions on the $\beta$-function, from the outset, was that those roots $u_s$ which are real satisfy $u_s<1$ [cf.~the comments after Eqs.~(\ref{roots})]. This means that for such $u_s$, the branching point $z_{\ast}^{(s)}$ where $F(z_{\ast})=a_0/u_s$ ($> a_0$) cannot be achieved at real $z_{\ast}^{(s)}=x_{\ast}^{(s)}$, i.e., also in such cases $z_{\ast}^{(s)}$ must have $y_{\ast}^{(s)} \not= 0$, and thus we can have also in such a case nonzero winding numbers ${\vec n} \not= {\vec 0}$.

Here, the procedure described in the previous Sec.~\ref{subs:Lpole} for ${\rm Im} \; {\cal G}_{\vec n}(a_{\ast})$ and  ${\rm Re} \; {\cal G}_{\vec n}(a_{\ast})$ [for $a_{\ast} \equiv F(x_{\ast})$ in the interval $0 < a_{\ast} < a_0$, and for $n_j, N_k, {\cal N}_k=0, \pm 1, \ldots,$], is now performed for ${\rm Im} \; {\cal K}_{\vec n}(a_{\ast}^{(s)}; u_s)$ and  ${\rm Re} \; {\cal K}_{\vec n}(a_{\ast}^{(s)}; u_s)$, again with $a_{\ast}^{(s)}=F(x_{\ast}^{(s)})$ in the interval $0 < a_{\ast}^{(s)} < a_0$ and  for $n_j^{(s)}, N_k^{(s)}, {\cal N}_k^{(s)}=0, \pm 1, \ldots$, but now also for each $u_s$ ($s=1,\ldots,N$). This means that Eq.~(\ref{Lbp}) for the branching points represents two real equations, in analogy with Eqs.~(\ref{LpReIm})-(\ref{ImReG}).
\be
{\rm Im} {\cal K}_{\vec n}(a_{\ast}^{(s)}; u_s) = 0, \quad
{\rm Re} {\cal K}_{\vec n}(a_{\ast}^{(s)}; u_s) = y_{\ast}^{(s)},    
\label{LbpReIm}
\ee
where
\bea
\lefteqn{
{\rm Im} {\cal K}_{\vec n}(a_{\ast}^{(s)}; u_s) \equiv   \frac{1}{\beta_0 a_0} {\Bigg \{}
\left( - u_s + \frac{a_0}{a_{\ast}^{(s)}} \right) +  B_0 \left[  - \ln_{(\rm pb)} (u_s - 1) + \ln \left( \frac{a_0}{a_{\ast}^{(s)}} - 1 \right) \right]
}
\nonumber\\
&&
+ \sum_{j=2 P + 1}^M B_j \left[  - \ln_{(\rm pb)} (u_s - t_j) + \ln \left( \frac{a_0}{a_{\ast}^{(s)}} - t_j \right) \right]
\nonumber\\ &&
+ \sum_{k=0}^{P-1} \frac{2
\left( ({\rm Re}B_{2 k+1}) ({\rm Re} t_{2 k +1}) - {\rm Re} (t_{2 k +1}^{\ast} B_{2 k +1}) \right)} {| {\rm Im}  t_{2 k +1} |}
\left[ - {\rm ArcTan} \left( \frac{ u_s - {\rm Re} t_{2 k +1}}{| {\rm Im}  t_{2k +1} |} \right) +
  {\rm ArcTan} \left( \frac{ a_0/a_{\ast}^{(s)} - {\rm Re} t_{2 k +1}}{| {\rm Im}  t_{2k +1} |} \right) \right]
\nonumber\\
&& + \sum_{k=0}^{P-1} ({\rm Re} B_{2 k +1}) \left[
- \ln \left( u_s^2 - 2 ({\rm Re} t_{2 k +1}) u_s + |t_{2 k +1}|^2  \right) +
\ln  \left( \left(\frac{a_0}{a_{\ast}^{(s)}}\right)^2 - 2 ({\rm Re} t_{2 k +1}) \frac{a_0}{a_{\ast}^{(s)}} + |t_{2 k +1}|^2 \right)
\right] {\Bigg \}}
\nonumber\\ &&
+ \frac{2 \pi}{\beta_0 a_0} \sum_{k=0}^{P-1} \frac{ \left( {\rm Re}(t_{2 k +1}^{\ast} B_{2 k +1}) - ({\rm Re} B_{2 k+1})({\rm Re} t_{2 k+1}) \right)}{| {\rm Im}  t_{2 k+ 1} |} {\cal N}_k^{(s)},
\label{ImK}
\eea
and
\bea
{\rm Re} {\cal K}_{\vec n}(a_{\ast}^{(s)}; u_s) & \equiv &
\frac{2 \pi}{\beta_0 a_0} \left\{
\left( B_0 n_0^{(s)} + \sum_{j=2 P +1}^M B_j n_j^{(s)} \right)
+ \sum_{k=0}^{P-1} ({\rm Re} B_{2 k +1}) N_k^{(s)} \right\}.
\label{ReK}
\eea 
We notice that ${\rm Im} {\cal K}_{\vec n}(a_{\ast}^{(s)}; u_s)$ depends only on the winding numbers ${\cal N}_k^{(s)}$, cf.~Eq.~(\ref{Ka}).
The existence of a Landau branching point means that Eqs.~(\ref{LbpReIm}) have a solution, for an $s$, and $a_{\ast}^{(s)}$ and $y_{\ast}^{(s)}$ such that: $0 < a_{\ast}^{(s)} < a_0$ and $|y_{\ast}^{(s)}| < \pi$.

The procedures described in this Section \ref{sec:sing}  for ${\rm Im} \; {\cal G}_{\vec n}(a_{\ast})$ and  ${\rm Re} \; {\cal G}_{\vec n}(a_{\ast})$, and for  ${\rm Im} \; {\cal K}_{\vec n}(a_{\ast}^{(s)}; u_s)$ and  ${\rm Re} \; {\cal K}_{\vec n}(a_{\ast}^{(s)}; u_s)$, represent a relatively simple algebraic instrument for practical verification of whether the pQCD scheme with a given beta function of the form (\ref{beta1}) described in Sec.~\ref{sec:impl} has Landau singularities or has no such singularities, and where these singularities are.

\section{Practical examples}
\label{sec:prac}

We will consider three specific cases of application of the above algebraic formalism: (a) when the $\beta(F)$ function (\ref{beta1}) has (cubic) polynomial structure and only real roots: $M=2$, $N=0$; $t_1, t_2 \in \mathbb{R}$; (b) $\beta(F)$ has (cubic) polynomial structure and complex roots:  $M=2$, $N=0$; $t_2=t_1^{\ast} \not\in \mathbb{R}$; (c) $\beta(F)$ has a Pad\'e structure with (one) pole: $M=N=1$. Although these are cases with low indices $M$ and $N$, we believe that they are representative to a certain degree, and show in practice how the presented formalism works. \textcolor{black}{The cases (a) and (b) are specific low-index cases belonging to the set of beta-functions discussed in Sec.~\ref{subs:Lpole} where Landau poles are expected to appear in the complex $Q^2$-plane. The case (c) is a specific low-index case belonging to the set of beta-functions discussed in Sec.~\ref{subs:Lbranch} where a cut structure of Landau singularities is expected.}

\subsection{Polynomial $\beta$ with real roots}
\label{subs:rr}

\begin{figure}[htb] 
\begin{minipage}[b]{.49\linewidth}
  \centering\includegraphics[width=85mm]{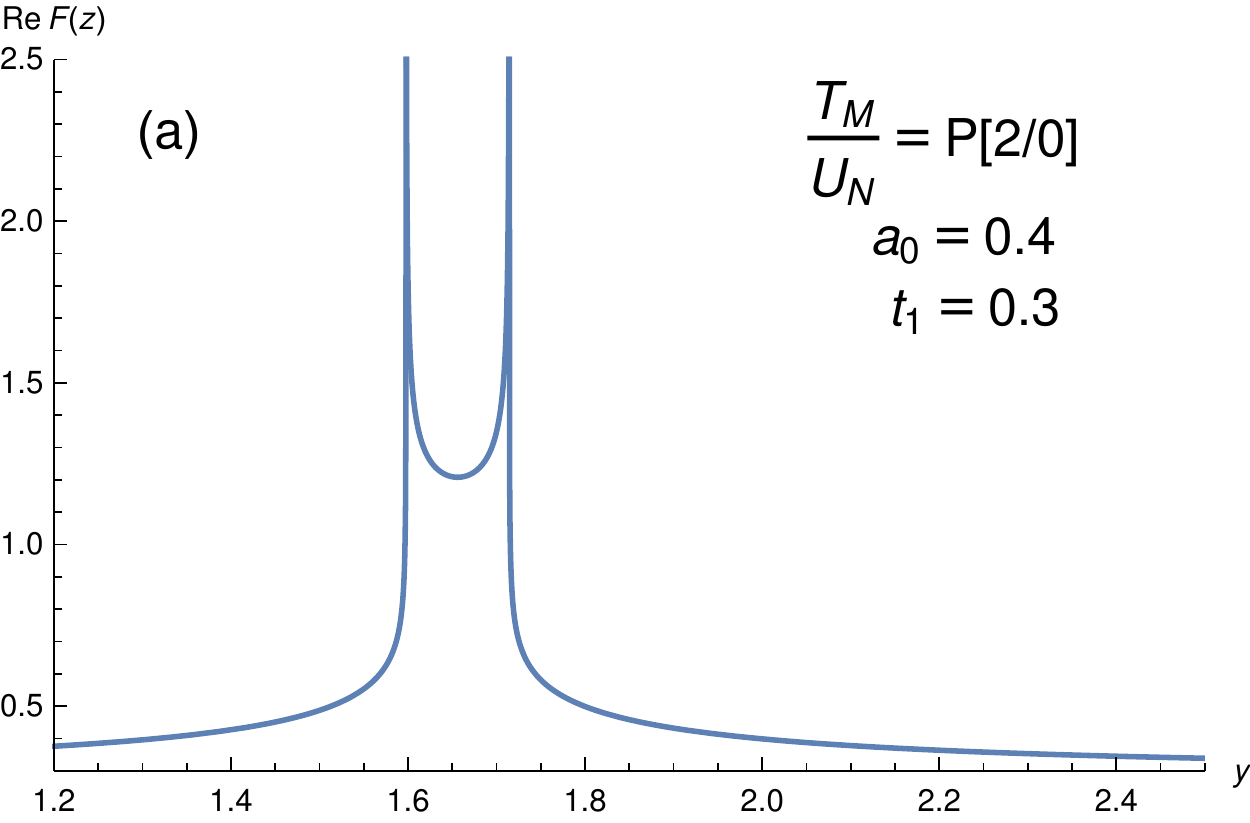}
  \end{minipage}
\begin{minipage}[b]{.49\linewidth}
  \centering\includegraphics[width=85mm]{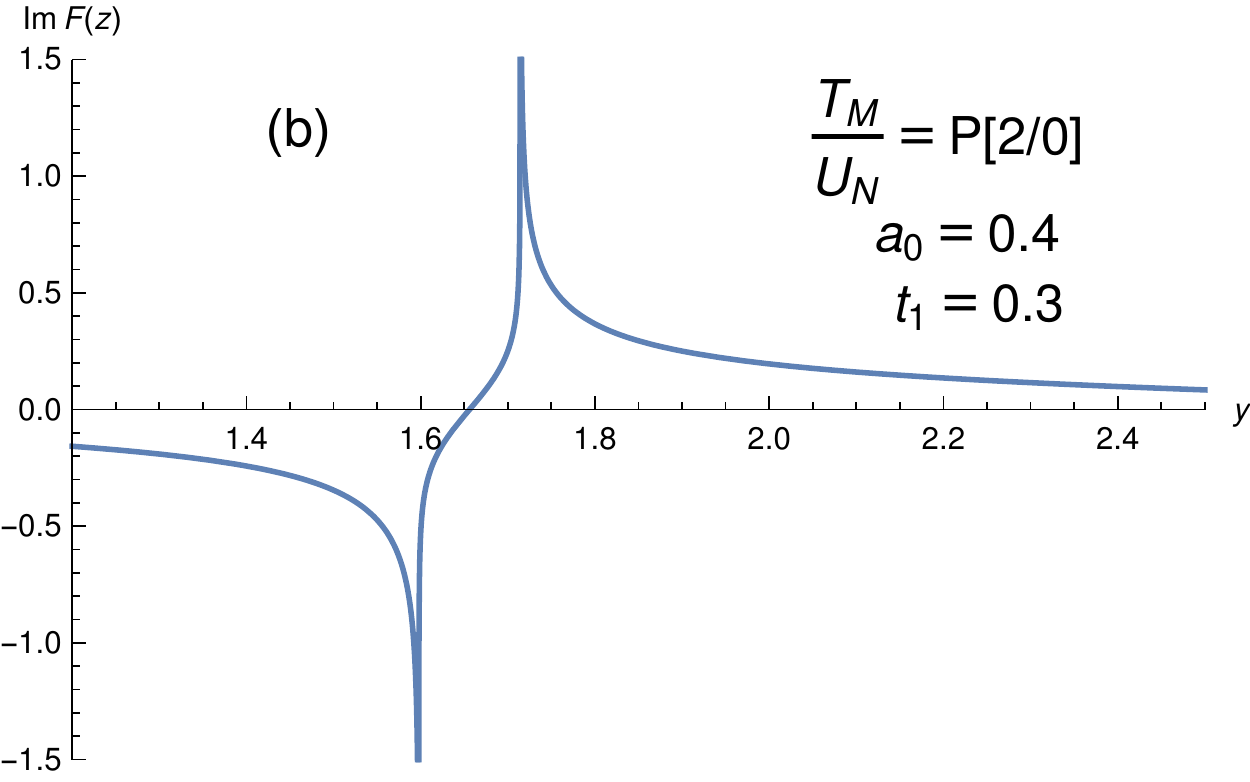}
\end{minipage}
\vspace{-0.2cm}
\caption{\footnotesize  (a) The real part of the running coupling $F(z)$ for $z=x_{\ast} + i y$ for $x_{\ast}=-5.03423$ and $0 < y \leq \pi$, for the considered case of Eqs.~(\ref{betarr})-(\ref{a0t1rr}); (b) the same as in (a), but for the imaginary part of $F(z)$.}
\label{FigReImarr}
\end{figure}

\begin{figure}[htb] 
\begin{minipage}[b]{.49\linewidth}
  \centering\includegraphics[width=85mm]{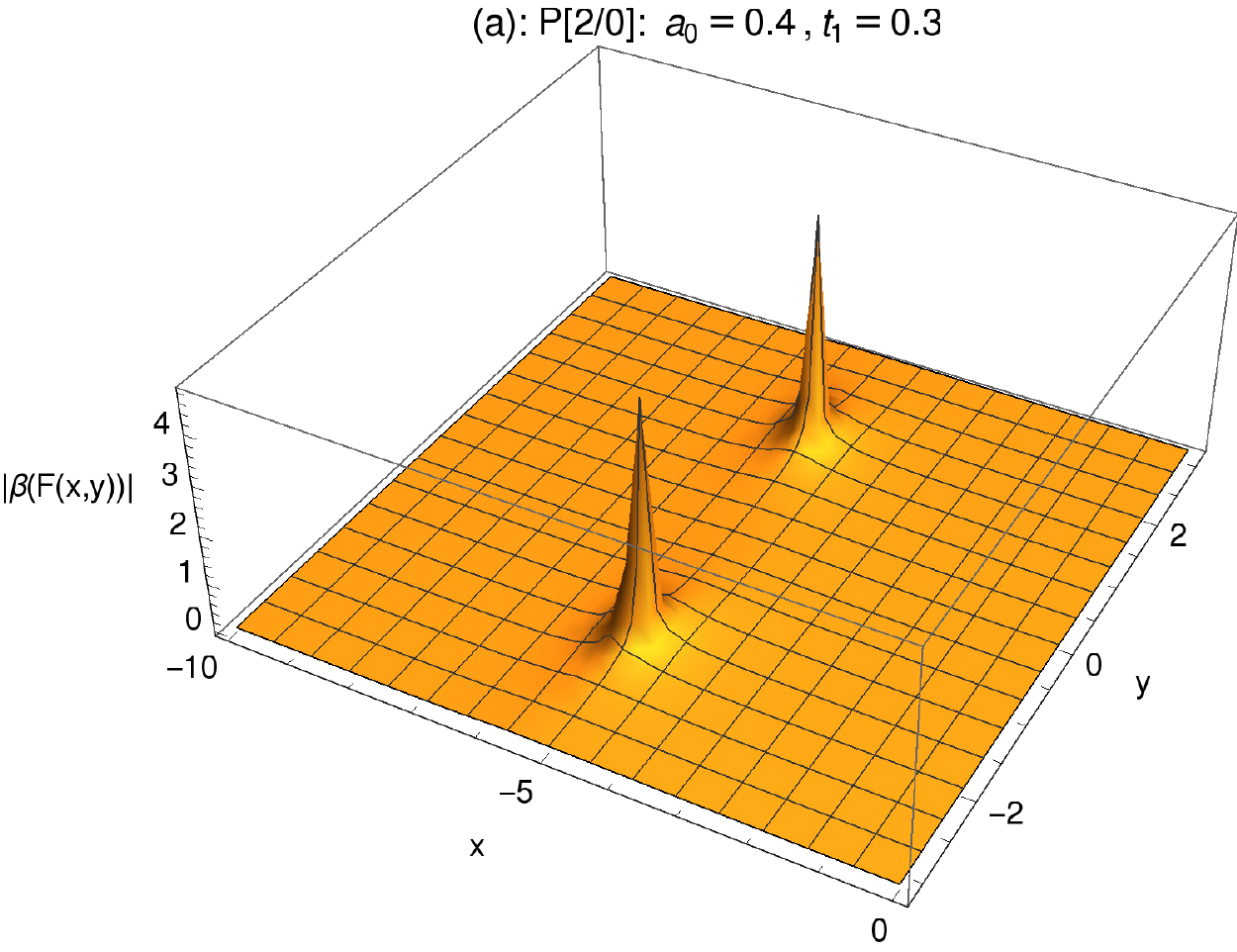}
  \end{minipage}
\begin{minipage}[b]{.49\linewidth}
  \centering\includegraphics[width=85mm]{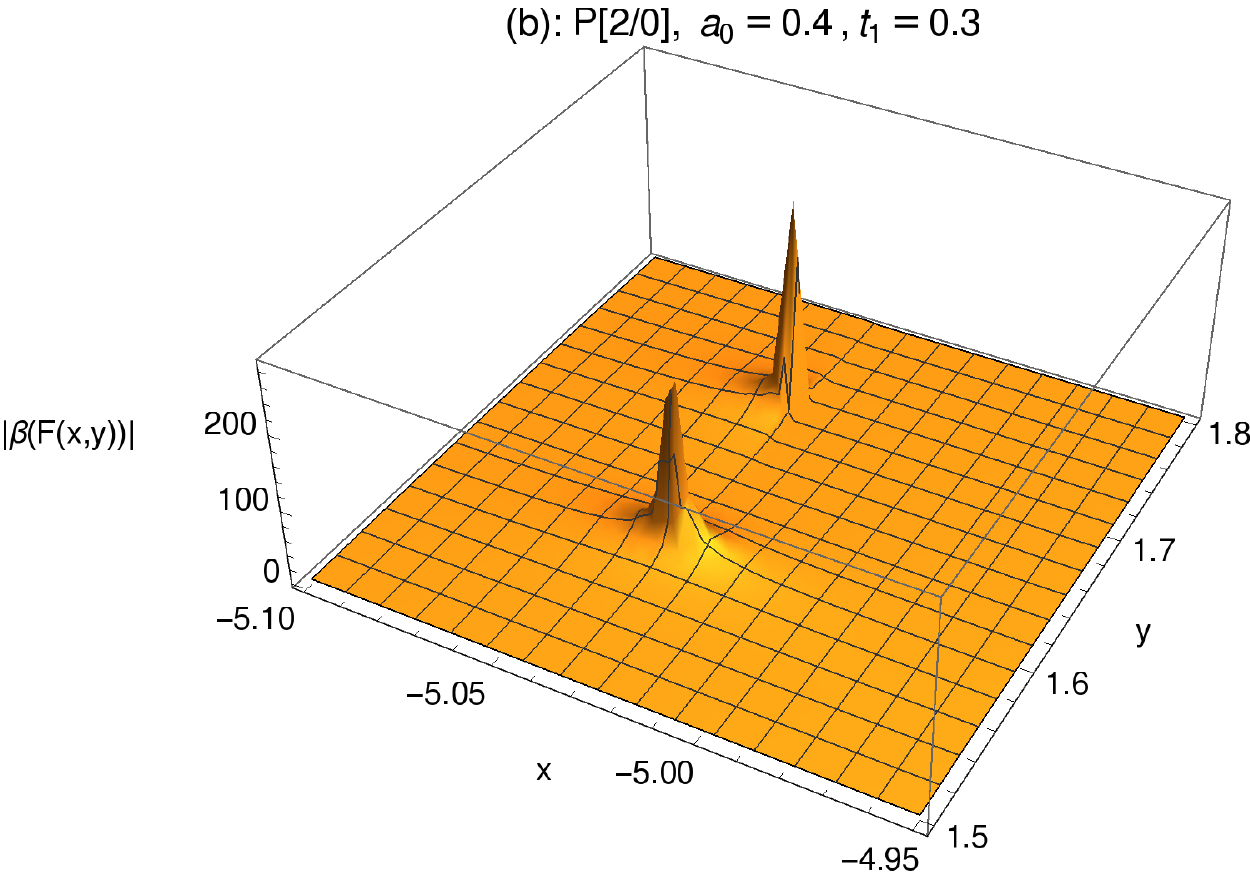}
\end{minipage}
\vspace{-0.2cm}
\caption{\footnotesize  (a) The numerical values of $|\beta(F(z))|$ in the the physical $z$-stripe  ($- \pi \leq y \equiv {\rm Im} z < \pi$), corresponding to the first Riemann sheet of the complex momenta $Q^2$. The numerical results indicate only one Landau pole in this region, and its complex conjugate. (b) ``Zoomed'' numerical calculation indicates two mutually close Landau poles in this region (and their complex conjugates). The calculation was performed using Mathematica software \cite{Mathematica}.}
\label{Figabsbetrr}
\end{figure}

Here we consider the case of ($M=2$, $N=0$)
\be
\beta(F) = - \beta_0 F^2 (1 - Y) \times P[2/0](Y){\big \vert}_{Y \equiv F/a_0} = - \beta_0 F^2 (1 - Y) (1 - t_1 Y) (1 - t_2 Y){\big \vert}_{Y \equiv F/a_0},
\label{betarr}
\ee
where $t_1$ and $t_2$ are real [and $t_1, t_2 < 1$ by physical requirements, cf.~the text after Eqs.~(\ref{roots})]. In order to present numerical results, we choose specific numerical input values for $a_0$ ($>0$) and $t_1$ (which we choose to be positive)
\be
a_0=0.4; \qquad t_1=+0.3.
\label{a0t1rr}
\ee
The condition (\ref{beta1con}) then gives
\be
t_2=-1 - (\beta_1/\beta_0) a_0 - t_1  \; ( \approx -2.0111),
\label{t2rr} \ee
where the numerical value is obtained by using in the universal $\beta$-coefficients $\beta_0$ and $\beta_1$ the number of active quark flavours $N_f=3$
($\beta_0=9/4$; $\beta_1=4$.) The resulting renormalization scheme parameters $c_j \equiv \beta_j/\beta_0$ ($j \geq 2$) are then $c_2=-14.4653$, $c_3=9.4271$ and $c_4=0$  (in $\MSbar$ scheme, for $N_f=3$, they are: $c_2=4.4711$, $c_3=20.990$, $c_4=56.588$).\footnote{When varying, at fixed $a_0$, the (real) $t_1$ ($t_1 < 1$), the (leading) scheme coefficient $c_2$ will vary according to the relation (\ref{cn}), and will be restricted in the range between  $-(1/a_0) (+3/a_0 + 2 c_1) < c_2 < (1/4) (1/a_0 +c_1) (-3/a_0 + c_1)$ [cf.~discussion just after Eqs.~(\ref{restr})], i.e., in our case of $a_0=0.4$ this is the range $-27.64 < c_2 < -6.12$.}
The $\kappa$ coefficient of Eqs.~(\ref{aQ2exp})-(\ref{kappa}) then has the value
\be
\kappa = \frac{\beta_0 a_0}{B_0} \approx 1.897.
\label{kapparr} \ee
Since $t_1$ and $t_2$ are real (hence: $M=2; P=0$), the only winding numbers (\ref{ndef}) are ${\vec n}=(n_0,n_1,n_2)$, and thus ${\rm Im} {\cal G}_{\vec n}(a_{\ast})$ is independent of ${\vec n}$. The first condition of Eq.~(\ref{LpReIm}) then immediately gives for $a_{\ast}$  (we recall: $0 < a_{\ast} < a_0$)
\be
{\rm Im} {\cal G}_{\vec n}(a_{\ast}) = 0 \; \Rightarrow  a_{\ast} \approx 0.320816;
\label{astrr} \ee
and the corresponding $x_{\ast}$ is\footnote{
We use throughout the reference value $\alpha_s(M_Z^2;\MSbar)=0.1179$ \cite{PDG2019}. This corresponds to the $N_f=3$ regime at $Q^2_{\rm in}= (2 {\bar m}_c)^2 = 2.54^2 \ {\rm GeV}^2$ to $a(Q^2_{\rm in}; \MSbar)$ ($ \equiv \alpha_s(Q^2_{\rm in};\MSbar)/\pi$) $=0.0834921$. We will use this reference value throughout (although, by using a different reference value is equivalent to changing the value of $Q^2_{\rm in}$ which does not affect our conclusions). The RGE-running from $M_Z^2$ down to $(2 {\bar m}_c)^2$ in $\MSbar$ is performed by using the five-loop RGE \cite{5lMSbarbeta} with four-loop quark threshold conditions at $\mu^2_{\rm thr.} = (2 {\bar m}_q)^2$ \cite{4lquarkthresh1,4lquarkthresh2}, where the $\MSbar$ quark mass values for ${\bar m}_q \equiv {\bar m}_q({\bar m}_q^2)$ was taken ${\bar m}_b=4.20$ GeV and ${\bar m}_c=1.27$ GeV. The transition from the (five-loop) $\MSbar$ scheme to the scheme of the considered $\beta$-function was performed at the scale $Q^2=(2 {\bar m}_c)^2$ and $N_f=3$, according to the approach as explained, e.g., in Ref.~\cite{3dAQCD} [Eq.~(13) there]. This gives, in the considered scheme of the $\beta$-function (\ref{betarr}), the value $a(Q^2_{\rm in})=0.0737597$.}
\be
x_{\ast}=-5.03423,
\label{xstrr} \ee
as can be easily checked by the implicit solution (\ref{implsolg}) when using there for $F(z)$ the value of $a_{\ast}$ Eq.~(\ref{astrr}).
When we now numerically integrate the RGE (\ref{RGE1}) along the line ${\rm Re}(z)=x_{\ast}$ in the $z$-plane,\footnote{This integration is 1-dimensional, much simpler and considerably more stable than the integration in the entire physical complex-$z$ stripe of Fig.~\ref{Qzplane}(b). We refer to this 1-dimensional integration as a seminumeric part of the procedure.} we obtain for the real and imaginary part of the coupling $F(z=x_{\ast}+ i y)$ the values presented in Figs.~\ref{FigReImarr},
which clearly show that there are singularities (poles) of the running coupling $a(Q^2) \equiv F(z)$ at $z_{\ast, \pm}^{(j)}=x_{\ast} \pm i y_{\ast}^{(j)}$
\be
y_{\ast}^{(1)} =1.59783;  \qquad y_{\ast}^{(2)} =1.71434.
\label{ystrr} \ee
The obtained points $z_{\ast, \pm}^{(j)}$ are the Landau poles. We can cross-check that these points are really the Landau poles by evaluating the algebraic expression ${\cal G}_{\vec n}(a_{\ast})$, Eq.~(\ref{Ga}), for various winding numbers ${\vec n}=(n_0,n_1,n_2)$, and we find that
\bes
\label{calGrr}
\bea
{\cal G}_{\vec n}(a_{\ast}) &=& +y_{\ast}^{(1)} (=+1.59783) \qquad {\rm for} \; {\vec n}=(0,0,0);
\label{calGrra} \\
{\cal G}_{\vec n}(a_{\ast}) &=& -y_{\ast}^{(1)} (=-1.59783) \qquad {\rm for} \; {\vec n}=(-1,-1,0);
\label{calGrrb} \\
{\cal G}_{\vec n}(a_{\ast}) &=& +y_{\ast}^{(2)} (=+1.71434) \qquad {\rm for} \; {\vec n}=(0,-1,0);
\label{calGrrc} \\
{\cal G}_{\vec n}(a_{\ast}) &=& -y_{\ast}^{(2)} (=-1.71434) \qquad {\rm for} \; {\vec n}=(-1,0,0);
\label{calGrrd}
\eea \ees
On the other hand, without the algebraic seminumeric approach described above, it would be difficult to find the (four) Landau poles on the first Riemann sheet of $Q^2$. In Figs.~\ref{Figabsbetrr}(a),(b) we present $|\beta(F(z))|$ for the considered couplings, obtained by the 2-dimensional numerical integration of the RGE in the complex $z$-stripe.  In Fig.~\ref{Figabsbetrr}(a) it is difficult to see that there are two Landau poles close to each other, at positive (and negative) values of ${\rm Im}(z)=y$; only the strongly ``zoomed'' Fig.~\ref{Figabsbetrr}(b) suggests that there are two poles near to each other, at $z_{\ast,+}^{(j)}=x_{\ast} + i y_{\ast}^{(j)}$ ($j=1,2$), as clearly obtained in Eqs.~(\ref{calGrr}) by our algebraic seminumeric analysis.

\subsection{Polynomial $\beta$ with complex roots}
\label{subs:cr}

Here we consider the case of ($M=2$, $N=0$)
\be
\beta(F) = - \beta_0 F^2 (1 - Y) \times P[2/0](Y){\big \vert}_{Y \equiv F/a_0} =- \beta_0 F^2 (1 - Y) (1 - t_1 Y) (1 - t_2 Y){\big \vert}_{Y \equiv F/a_0},
\label{betacr}
\ee
where $t_1$ and $t_2$ are complex nonreal and thus mutually complex conjugate [$t_2 = (t_1)^{\ast}$]. Since we want to present numerical results, we choose as an example the following specific input values:
\be
a_0=0.5; \qquad {\rm Im} t_1=+0.60.
\label{a0t1cr}
\ee
The condition (\ref{beta1con}) then gives
\be
{\rm Re} t_1=- \frac{1}{2} \left( 1 + (\beta_1/\beta_0) a_0 \right)  \; ( \approx -0.9444),
\label{Ret1} \ee
where, as in Sec.~\ref{subs:rr},  the numerical value is obtained by using in the universal $\beta$-coefficients $\beta_0$ and $\beta_1$ with $N_f=3$
The renormalization scheme parameters $c_j \equiv \beta_j/\beta_0$ ($j \geq 2$) are in this case $c_2=-2.5477$, $c_3=-10.0158$ and $c_4=0$.\footnote{When varying, at fixed $a_0$, the complex $t_1$ in this case (here ${\rm Im} t_1$ can be regarded as the only free parameter), the (leading) scheme coefficient $c_2$ will vary according to the relation (\ref{cn}), and will be restricted in the range between  $(1/4) (1/a_0 +c_1) (-3/a_0 + c_1) \leq c_2$ [cf.~discussion just after Eqs.~(\ref{restr})], i.e., in our case of $a_0=0.5$ this is the range $-3.988 \leq c_2$.}
The coefficient $\kappa$ of Eqs.~(\ref{aQ2exp})-(\ref{kappa}), has now the value
\be
\kappa = \frac{\beta_0 a_0}{B_0} \approx 4.6585
\label{kappacr} \ee
Since $t_1$ and $t_2$ are complex nonreal (hence: $M=2; P=1$), the winding numbers (\ref{ndef}) are ${\vec n}=(n_0,N_0,{\cal N}_0)$, and thus ${\rm Im} {\cal G}_{\vec n}(a_{\ast})$ depends on ${\cal N}_0$ and ${\rm Re} {\cal G}_{\vec n}(a_{\ast})$ depends on $n_0$ and $N_0$.

\begin{figure}[htb] 
\begin{minipage}[b]{.49\linewidth}
  \centering\includegraphics[width=85mm]{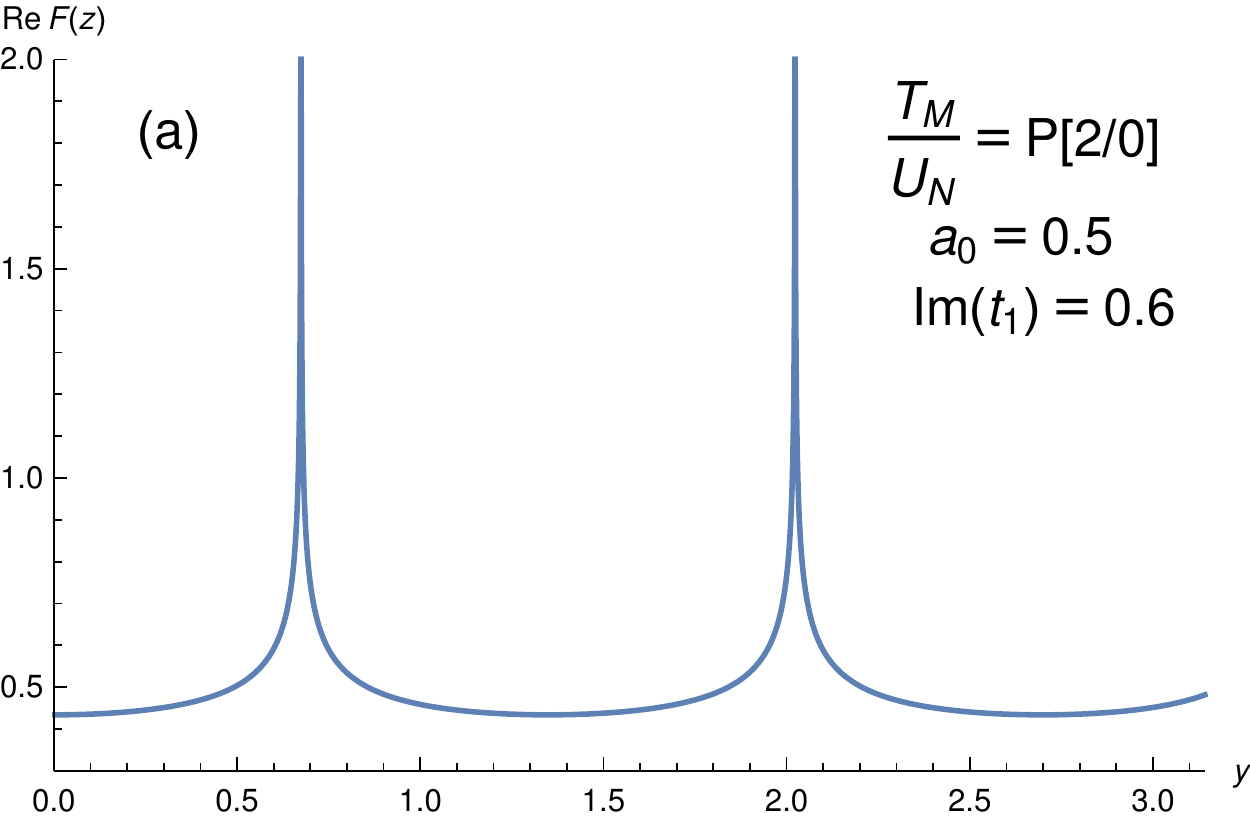}
  \end{minipage}
\begin{minipage}[b]{.49\linewidth}
  \centering\includegraphics[width=85mm]{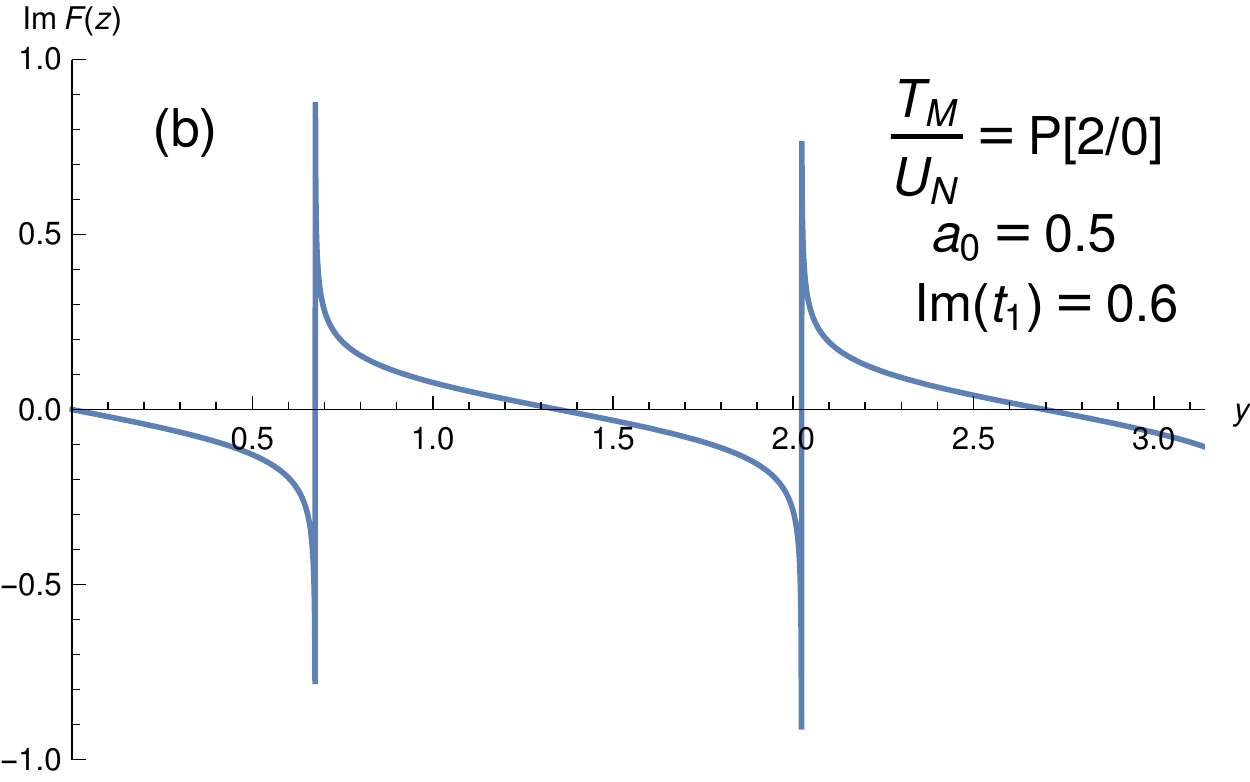}
\end{minipage}
\vspace{-0.2cm}
\caption{\footnotesize  (a) The real part of the running coupling $F(z)$ for $z=x_{\ast} + i y$ for $x_{\ast}=-4.18465$ and $0 < y \leq \pi$, for the considered case of Eqs.~(\ref{betacr})-(\ref{a0t1cr}); (b) the same as in (a), but for the imaginary part of $F(z)$.}
\label{FigReImacr}
\end{figure}

\begin{figure}[htb] 
\begin{minipage}[b]{.49\linewidth}
  \centering\includegraphics[width=85mm]{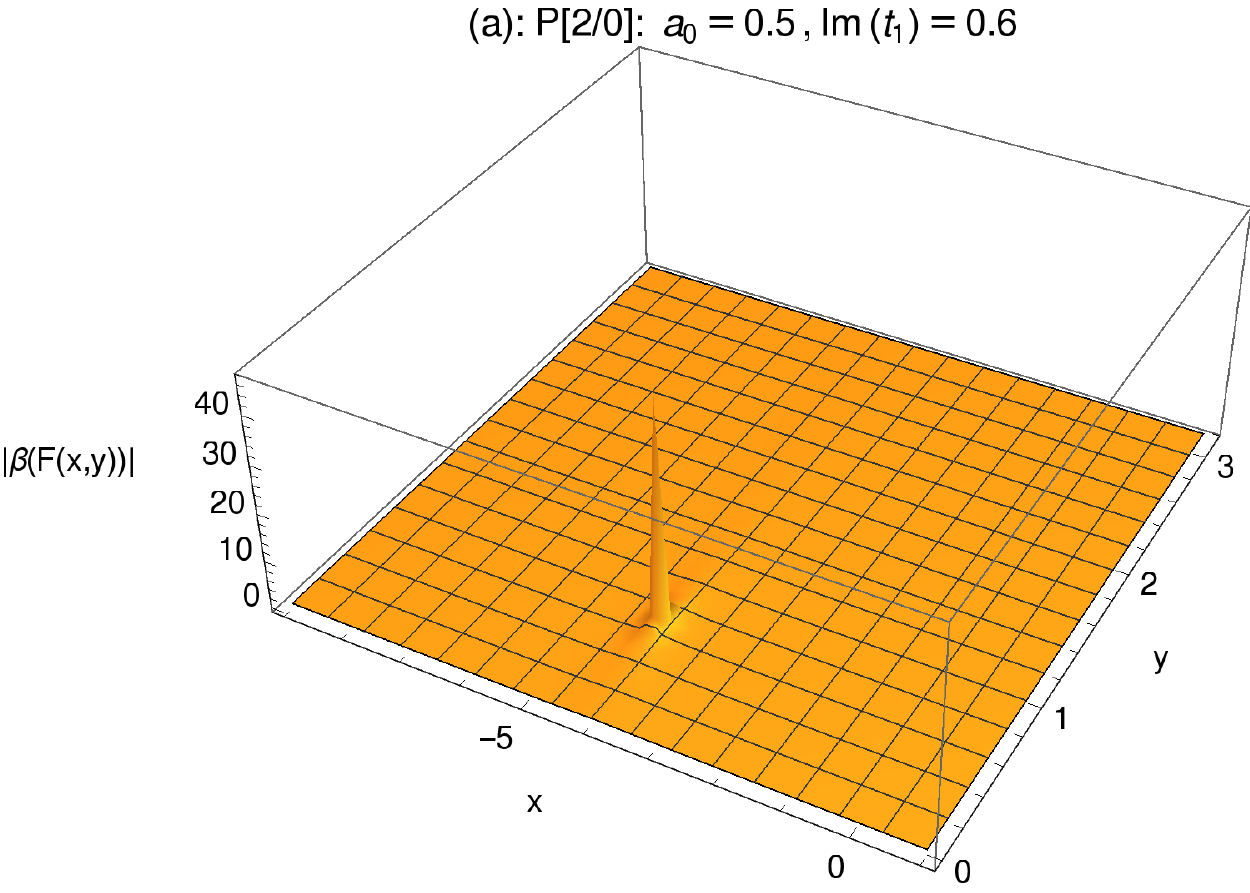}
  \end{minipage}
\begin{minipage}[b]{.49\linewidth}
  \centering\includegraphics[width=85mm]{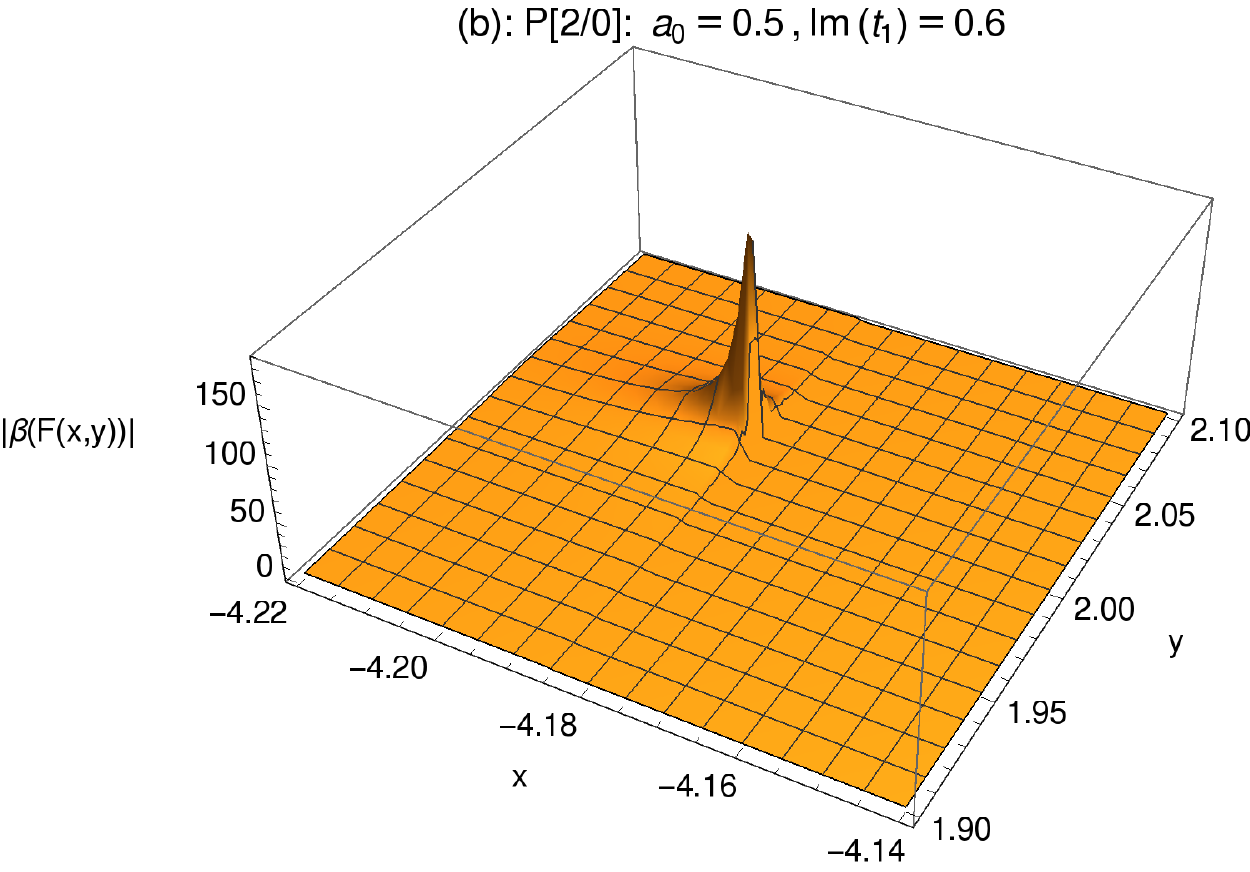}
\end{minipage}
\vspace{-0.2cm}
\caption{\footnotesize  (a) The numerical values of $|\beta(F(z))|$ in the upper half of the physical $z$-stripe   ($0 \leq y \equiv {\rm Im} z < \pi$), corresponding to the upper half of the first Riemann sheet of the complex momenta $Q^2$. The numerical results suggest the existence of a Landau pole at $z=x_{\ast} + i \; 0.67438$  where $x_{\ast}=-4.18465$ (and at its complex conjugate  $z=x_{\ast} - i \; 0.67438$). (b) ``Zoomed'' numerical calculation indicates the existence of an additional Landau pole at $z=x_{\ast} + i \; 2.02315$ (and its complex conjugate).}
\label{Figabsbetcr}
\end{figure}

The first condition of Eq.~(\ref{LpReIm}) then gives for $a_{\ast}$  the acceptable solution (i.e., in the interval $0 < a_{\ast} < a_0$) only when ${\cal N}_0 \geq -1$
\bes
\bea
 {\rm Im} {\cal G}_{\vec n}(a_{\ast}) &=& 0 \; \Rightarrow
 \\
 a_{\ast} &\approx & 0.492229({\cal N}_0=-1); \; 0.433899({\cal N}_0=0); \; 0.305849({\cal N}_0=1); \; 0.215326({\cal N}_0=2); \; {\rm etc.}
\label{astcr} \eea \ees
and the corresponding $x_{\ast}$ (we use $\alpha_s(M_Z^2;\MSbar)=0.1179$ as described in Sec.~\ref{subs:rr}) is obtained from the implicit solution (\ref{implsolg}) with $F(z)=a_{\ast}$
\be
x_{\ast}=-4.73364({\cal N}_0=-1); \; -4.18465({\cal N}_0=0); \; -3.63565({\cal N}_0=1); \; -3.08666 ({\cal N}_0=2); \; {\rm etc.}
\label{xstcr} \ee
When we now perform the simple (1-dimensional) numerical integration of the RGE (\ref{RGE1}) along the line ${\rm Re}(z)=x_{\ast}$ in the $z$-plane, we obtain for the real and imaginary part of the coupling $F(z=x_{\ast}+ i y)$ on the first Riemann sheet ($|y| \leq \pi$) singular structure only when ${\cal N}_0=0$ ($x_{\ast}=-4.18465$), with the values presented in Figs.~\ref{FigReImacr}.
These Figures clearly show that there are singularities (poles) of the running coupling $a(Q^2) \equiv F(z)$ at $z_{\ast, \pm}^{(j)}=x_{\ast} \pm i y_{\ast}^{(j)}$
\be
y_{\ast}^{(1)} =0.67438;  \qquad y_{\ast}^{(2)} =2.02315.
\label{ystcr} \ee
As in Sec.~\ref{subs:rr}, we conclude that the obtained points $z_{\ast, \pm}^{(j)}$ are the Landau poles. We cross-check that these points are really the Landau poles by evaluating the algebraic expression ${\cal G}_{\vec n}(a_{\ast})$, Eq.~(\ref{Ga}),  for various values of the winding numbers ${\vec n}=(n_0,N_0,{\cal N}_0)$, and we find
\bes
\label{calGcr}
\bea
{\cal G}_{\vec n}(a_{\ast}) &=& +y_{\ast}^{(1)} (=+0.67438) \qquad {\rm for} \; {\vec n}=(0,0,0);
\label{calGcra} \\
{\cal G}_{\vec n}(a_{\ast}) &=& -y_{\ast}^{(1)} (=-0.67438) \qquad {\rm for} \; {\vec n}=(-1,0,0);
\label{calGcrb} \\
{\cal G}_{\vec n}(a_{\ast}) &=& +y_{\ast}^{(2)} (=+2.02315) \qquad {\rm for} \; {\vec n}=(+1,0,0);
\label{calGcrc} \\
{\cal G}_{\vec n}(a_{\ast}) &=& -y_{\ast}^{(2)} (=-2.02315) \qquad {\rm for} \; {\vec n}=(-2,0,0);
\label{calGcrd}
\eea \ees
On the other hand, the fully numerical (2-dimensional) integration of the RGE (\ref{RGE1}) in the first Riemann sheet of the complex squared momenta $Q^2$ (i.e., in the complex $z$-stripe with $|{\rm Im} z| \leq \pi$) gives us the results in Figs.~\ref{Figabsbetcr}(a),(b) where we present $|\beta(F(z))|$ for the considered couplings. In Fig.~\ref{Figabsbetcr}(a) it is hard to see two of the four mentioned Landau poles, namely those with ${\rm Im}(z)=\pm 2.02315$. Only the strongly ``zoomed'' Fig.~\ref{Figabsbetcr}(b) suggests that there are Landau poles also at $z=x_{\ast} \pm i \; 2.02315$.

\subsection{Pad\'e $\beta$ with a real pole}
\label{subs:Pa}
\begin{figure}[htb] 
\begin{minipage}[b]{.49\linewidth}
  \centering\includegraphics[width=85mm]{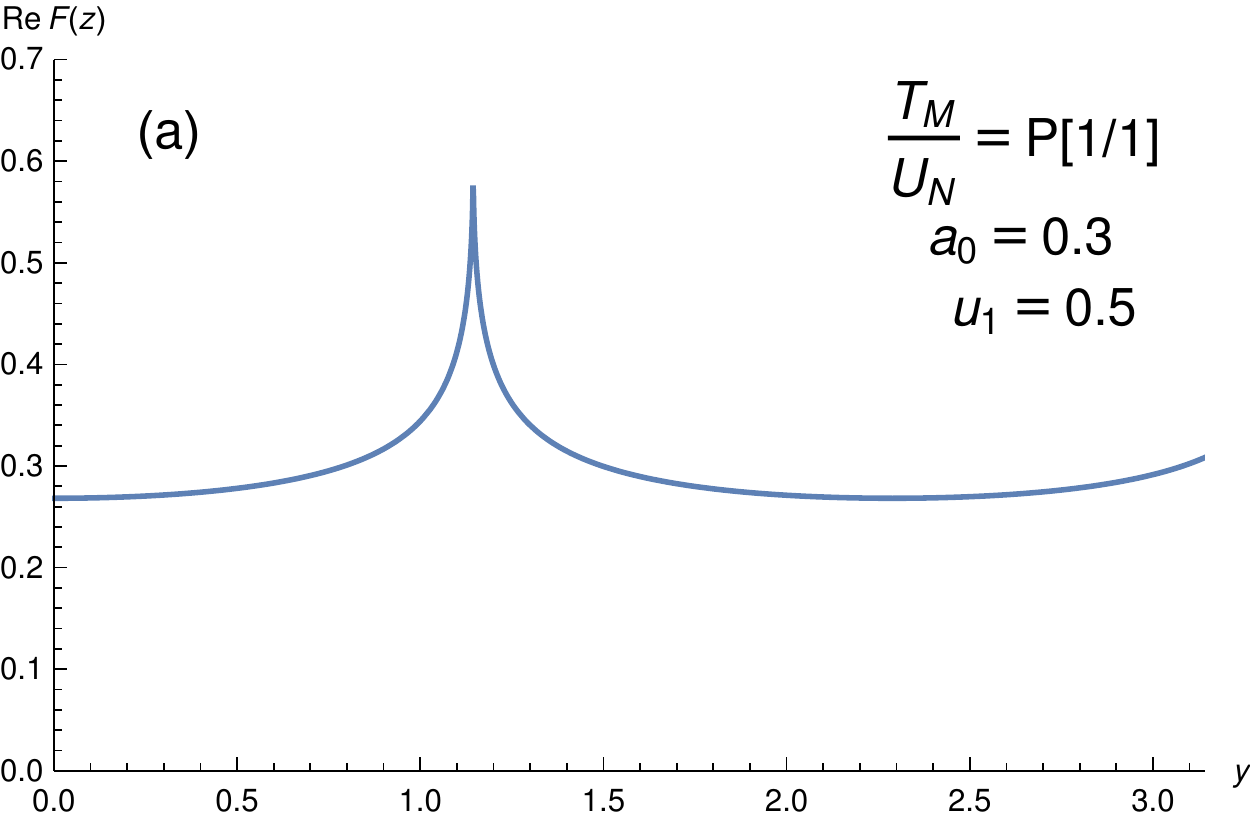}
  \end{minipage}
\begin{minipage}[b]{.49\linewidth}
  \centering\includegraphics[width=85mm]{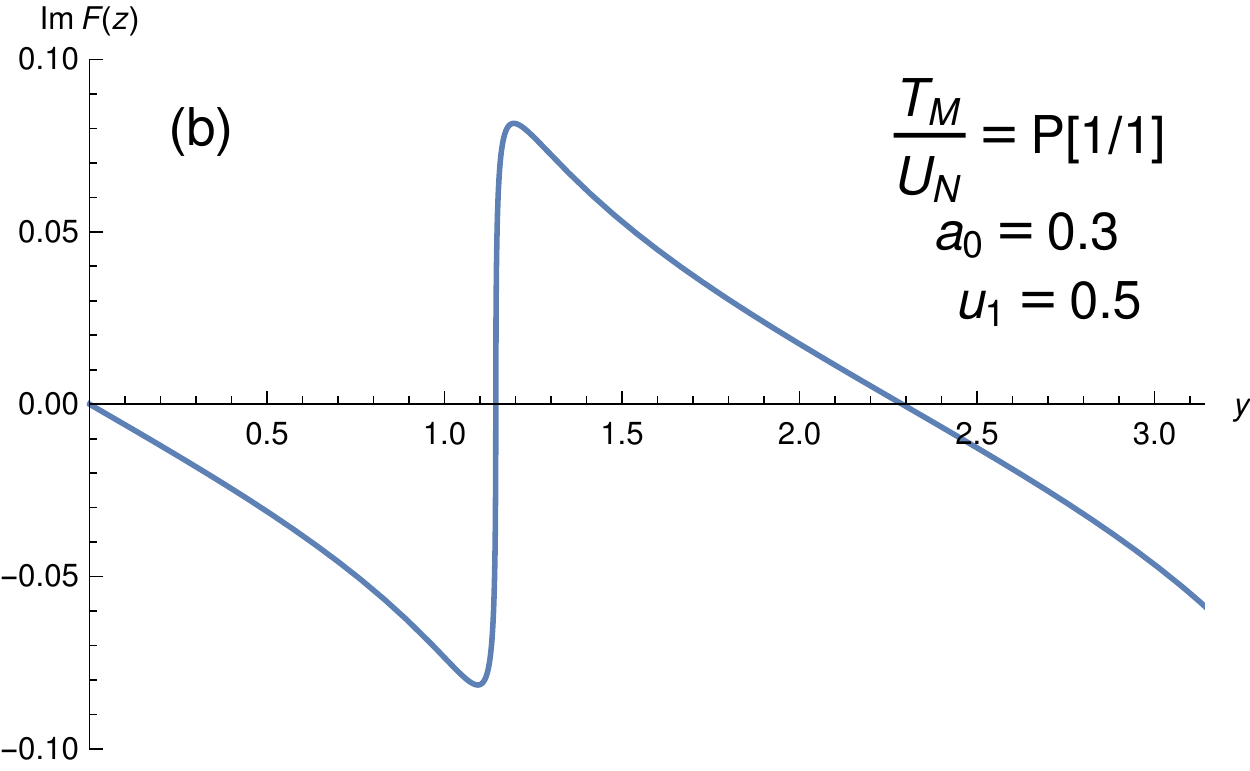}
\end{minipage}
\vspace{-0.2cm}
\caption{\footnotesize  (a) The real part of the running coupling $F(z)$ for $z=x_{\ast}^{(1)} + i y$ for $x_{\ast}^{(1)}=-4.38168$ and $0 < y \leq \pi$, for the considered case of Eqs.~(\ref{betaPa})-(\ref{a0u1Pa}); (b) the same as in (a), but for the imaginary part of $F(z)$.}
\label{FigReImaPa}
\end{figure}

Here we consider a numerical example for the types of $\beta$-function of Sec.~\ref{subs:Lbranch} where a finite-valued Landau branching point is realized. We will take the simplest case $M=1$ and $N=1$ (and $P=0$)  where $\beta$-function Eq.~(\ref{beta1}) has a Pad\'e form with one real pole
\be
\beta(F) = - \beta_0 F^2 (1 - Y) \times P[1/1](Y){\big \vert}_{Y \equiv F/a_0} = - \beta_0 F^2 (1 - Y) \frac{(1 - t_1 Y)}{(1 - u_1 Y)} {\bigg \vert}_{Y \equiv F/a_0}
\label{betaPa}
\ee
Here, both $u_1$ and $t_1$ are real and related via the relation (\ref{beta1con}). The $\beta$-function has a pole at the coupling value $F(z_{\ast}) = a_0/u_1$. We will present numerical results, so we choose as a representative example the following specific input values:
\be
a_0=0.3; \qquad u_1=+0.50.
\label{a0u1Pa}
\ee
The condition (\ref{beta1con}) then gives
\be
t_1=u_1 - \left( 1 + (\beta_1/\beta_0) a_0 \right) = 0.5 - \left( 1 + (\beta_1/\beta_0) a_0 \right)  \; ( \approx -1.0333),
\label{t1Pa} \ee
where, as in the previous examples, we use the values of $\beta_0$ and $\beta_1$ with $N_f=3$.
The resulting renormalization scheme parameters $c_j \equiv \beta_j/\beta_0$ ($j \geq 2$) are then $c_2=-8.5185$, $c_3=-14.1975$, $c_4=-23.6626$, etc.\footnote{When varying, at fixed $a_0$, the (real) $u_1$ ($u_1 < 1$ and $t_1<1$; $u_1$ is the only free degree of freedom), the (leading) scheme coefficient $c_2$ will vary according to the relation (\ref{cn}), and will be restricted according to Eq.~(\ref{MN11}).}
  The $\kappa$ coefficient of Eqs.~(\ref{aQ2exp})-(\ref{kappa}) has in this case the value
\be
\kappa = \frac{\beta_0 a_0}{B_0} = 2.7450.
\label{kappaPa} \ee
Since in the considered case we have $M=1$ and $P=0$ (and $N=1$), the only winding numbers are ${\vec n} = \{ n_0^{(1)}, n_1^{(1)} \}$.
The (real) value of the coupling $F(x^{(1)}_{\ast}) =a_{\ast}^{(1)}$ ($0 <a_{\ast}^{(1)} < a_0$)  is then obtained by the condition ${\rm Im} {\cal K}_{\vec n}(a_{\ast}^{(1)}, u_1)=0$, cf.~Eq.~(\ref{LbpReIm}), where ${\rm Im} {\cal K}_{\vec n}(a_{\ast}^{(1)}, u_1)$ is independent of the winding numbers ${\vec n} \equiv \{ n_0^{(1)}, n_1^{(1)} \}$. This then immediately gives us
\be
{\rm Im} {\cal K}(a_{\ast}^{(1)}, u_1) = 0 \; \Rightarrow a_{\ast}^{(1)}=0.268253.
\label{astPa} \ee
The corresponding value of $x^{(1)}_{\ast}$ is [we use $\alpha_s(M_Z^2;\MSbar)=0.1179$ as in Sec.~\ref{subs:rr}] is obtained from the implicit solution (\ref{implsolg}) with $F(z)=a_{\ast}$ at $z=x^{(1)}_{\ast}$
\be
x^{(1)}_{\ast} = -4.38168.
\label{xstPa} \ee
Now performing the simple (1-dimensional) numerical integration of the RGE (\ref{RGE1}) along the line ${\rm Re}(z)=x_{\ast}^{(1)}$ in the $z$-plane, gives us the real and imaginary part of the coupling $F(z=x_{\ast}^{(1)}+ i y)$ on the first Riemann sheet ($|y| \leq \pi$) with the values presented in Figs.~\ref{FigReImaPa}.
 These Figures clearly show that, for ${\rm Re}(z)=x_{\ast}^{(1)}$ ($=-4.38168$),  there is a singular behaviour of the running coupling $a(Q^2) \equiv F(z)$ in the Riemann sheet only at the points $z_{\ast, \pm}^{(j)}=x_{\ast}^{(1)} \pm i y_{\ast}^{(1)}$ where
\be
y_{\ast}^{(1)} \approx 1.14448.
\label{ystPa} \ee
On the other hand, the second condition in Eq.~(\ref{LbpReIm}) should give us in this case the same values $\pm y_{\ast}^{(1)} = \pm 1.14448$. Indeed, the evaluation of the algebraic expression ${\cal K}_{\vec n}(a_{\ast}^{(1)}, u_1)$, Eq.~(\ref{Ka}), gives
\be
{\cal K}_{\{0,0 \}}(a_{\ast}^{(1)}, u_1) \approx 1.14448, \qquad
{\cal K}_{\{-1,0\}}(a_{\ast}^{(1)}, u_1) \approx -1.14448.
\label{ReKPa} \ee
This is consistent with the results (\ref{ystPa}), and clearly shows that in the considered case the Landau branching point is achieved in the first Riemann sheet only at the two complex conjugate points $z_{\ast}^{(1)}=x_{\ast}^{(1)} \pm i  y_{\ast}^{(1)}$ with $x_{\ast}^{(1)}=-4.38168$ and $y_{\ast}^{(1)}=1.14448$,  and with the corresponding winding numbers $\{ {\vec n} \} \equiv \{ n_0^{(1)}, n_1^{(1)} \}$ equal to $\{ 0, 0, \}$ and $\{ -1, 0 \}$, respectively. Further, Figs.~(\ref{FigReImaPa}) indicate that the coupling $F(z_{\ast}^{(1)})$ at this point achieves the (real) value $0.60$ which coincides with the value $a_0/u_1$, i.e., the value where $\beta$-function diverges (but not the coupling).

\begin{figure}[htb] 
\begin{minipage}[b]{.49\linewidth}
 \centering\includegraphics[width=85mm]{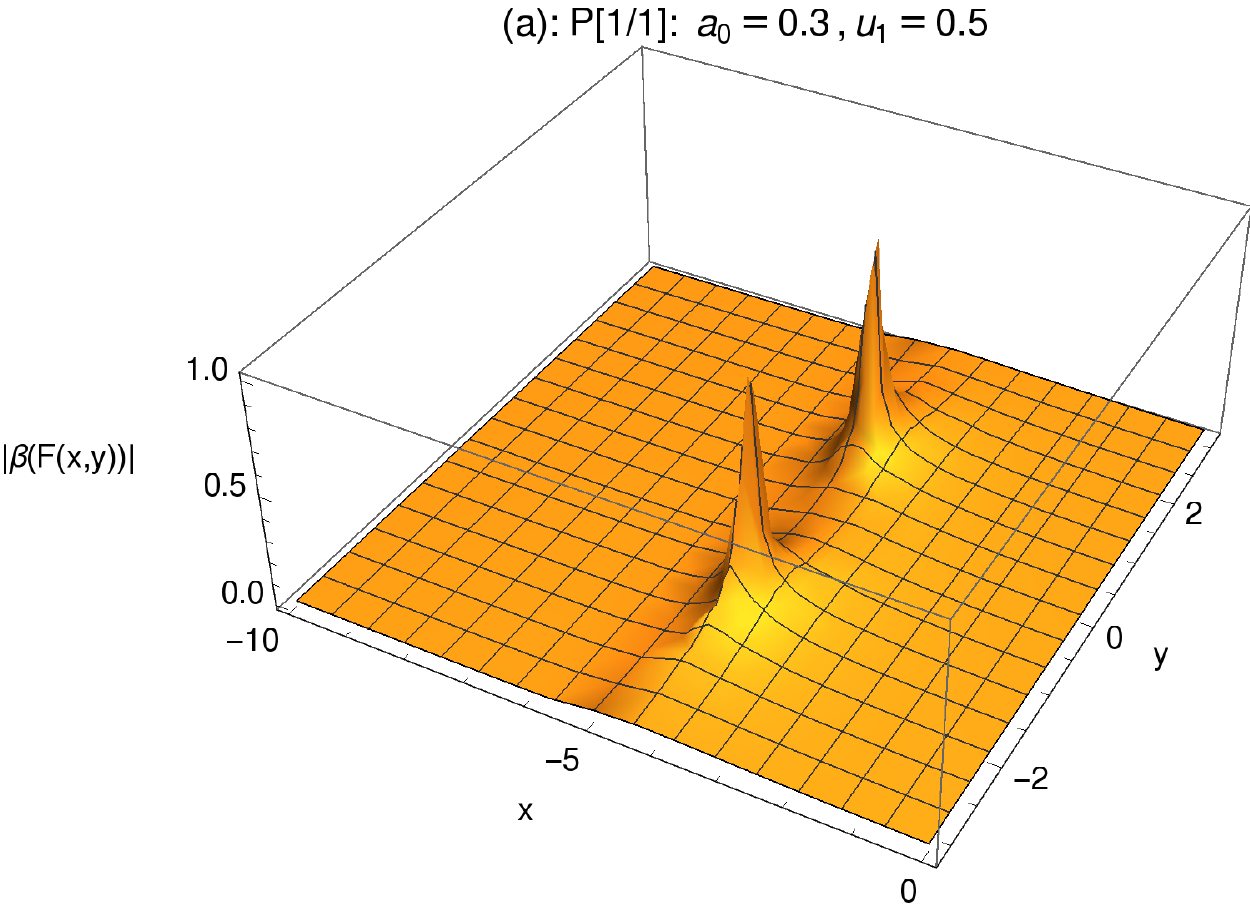}
  \end{minipage}
\begin{minipage}[b]{.49\linewidth}
 \centering\includegraphics[width=85mm]{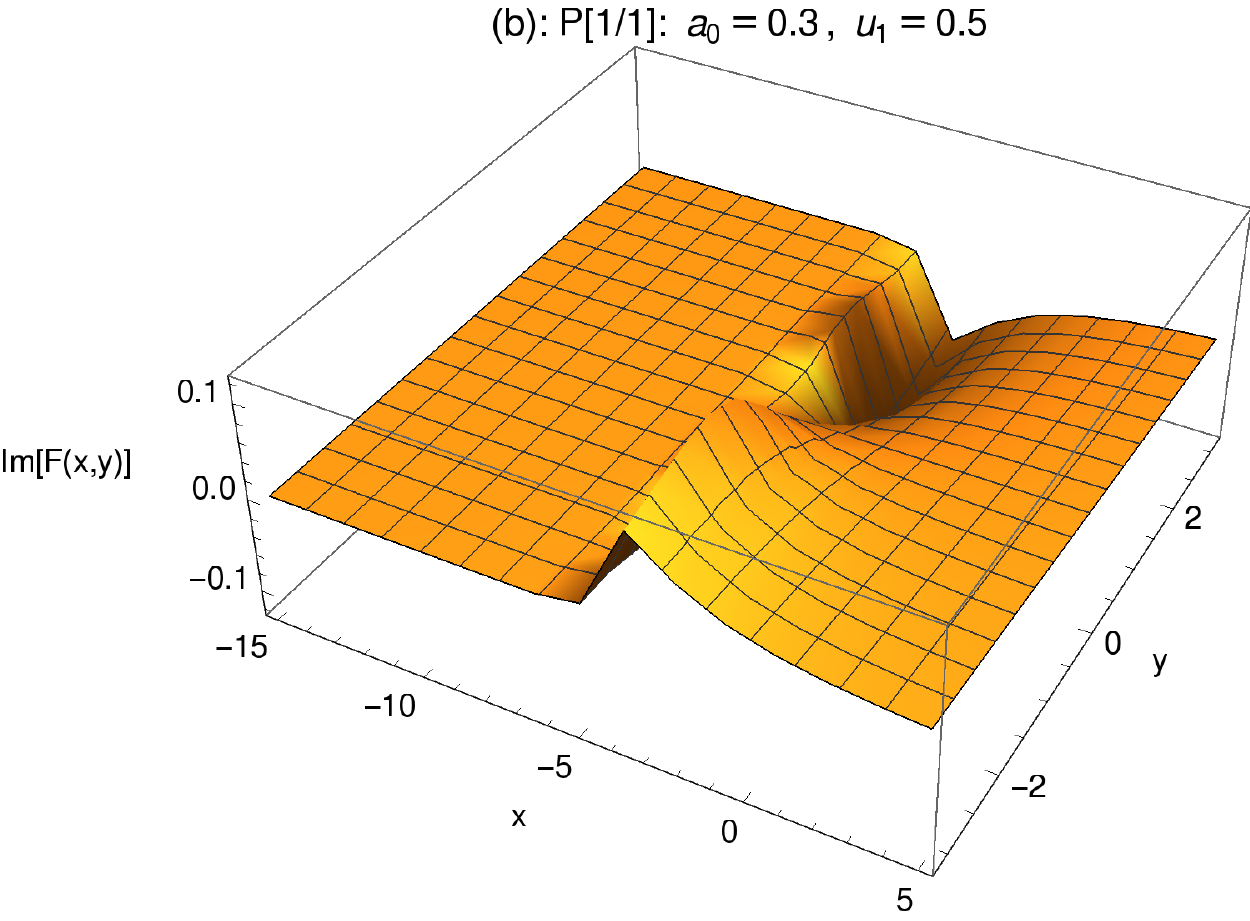}
\end{minipage}
\vspace{-0.2cm}
\caption{\footnotesize  (a) The numerical values of $|\beta(F(z))|$ in the physical $z$-stripe   ($- \pi \leq y \equiv {\rm Im} z < \pi$; $z=x + i y$), corresponding to the first Riemann sheet of the complex momenta $Q^2$. The numerical results indicate a complex conjugate pair $z=-4.382 \pm i \; 1.144$ for the Landau singularities. (b) The numerical values of ${\rm Im} F(z)$.}
\label{FigsPa}
 \end{figure}

The fully numerical (two-dimensional) integration of the RGE (\ref{RGE1}) in the first Riemann sheet of the complex squared momenta gives us the results in Figs.~\ref{FigsPa}.
Figure \ref{FigsPa}(a) shows $|\beta(F(z))|$ and indicates the Landau singularities at $z_{\ast,\pm}^{(1)}=x_{\ast}^{(1)} \pm i  y_{\ast}^{(1)}$. Figure \ref{FigsPa}(b) shows ${\rm Im} F(z)$ and indicates that the previously mentioned singularities are indeed branching points, with the cut in the complex-$z$ stripe extending from  $z_{\ast,+}^{(1)}=x_{\ast}^{(1)} + i  y_{\ast}^{(1)}$ along the line  $z =x_{\ast}^{(1)} + i  y$ with $y \geq y_{\ast}^{(1)}$, and the complex-conjugate cut from $z_{\ast,-}^{(1)}=x_{\ast}^{(1)} - i  y_{\ast}^{(1)}$ along the line $z =x_{\ast}^{(1)} + i  y$ with $y \leq -y_{\ast}^{(1)}$; the same indication can be obtained when evaluating ${\rm Re} F(z)$ in the $z$-complex stripe.\footnote{
In practice, the 2-dimensional numerical integration of the RGE in the $z$-complex stripe $|{\rm Im}z| \leq \pi$ [corresponding to the first Riemann sheet in the squared momentum plane $Q^2$ ($=-q^2 = Q^2_{\rm in} \exp(z)$)] was always performed first along the entire real $z$ axis, and then at each fixed real value of $z=x$ the RGE was integrated along the imaginary ($y$) direction of $z=x+i y$ ($-\pi \leq y < + \pi$).} For example, at $z=x^{(1)}_{\ast} + i 1.5$ we have numerically: $F(x^{(1)}_{\ast} + \epsilon +i 1.5) - F(x^{(1)}_{\ast} +i 1.5) \approx -0.189- i \; 0.327$ (when $\epsilon \approx 10^{-5}$-$10^{-3}$).
 
On the other hand, the algebraic seminumeric analysis above, Eqs.~(\ref{astPa})-(\ref{ReKPa}) and Figs.~\ref{FigReImaPa}, shows that the Landau singularities $z_{\ast,\pm}^{(1)}=x_{\ast}^{(1)} \pm i  y_{\ast}^{(1)}$ are indeed branching points (with cuts) and correspond to specific winding numbers, and that no other branching points exist in the first Riemann sheet.

\begin{figure}[htb] 
\centering\includegraphics[width=110mm]{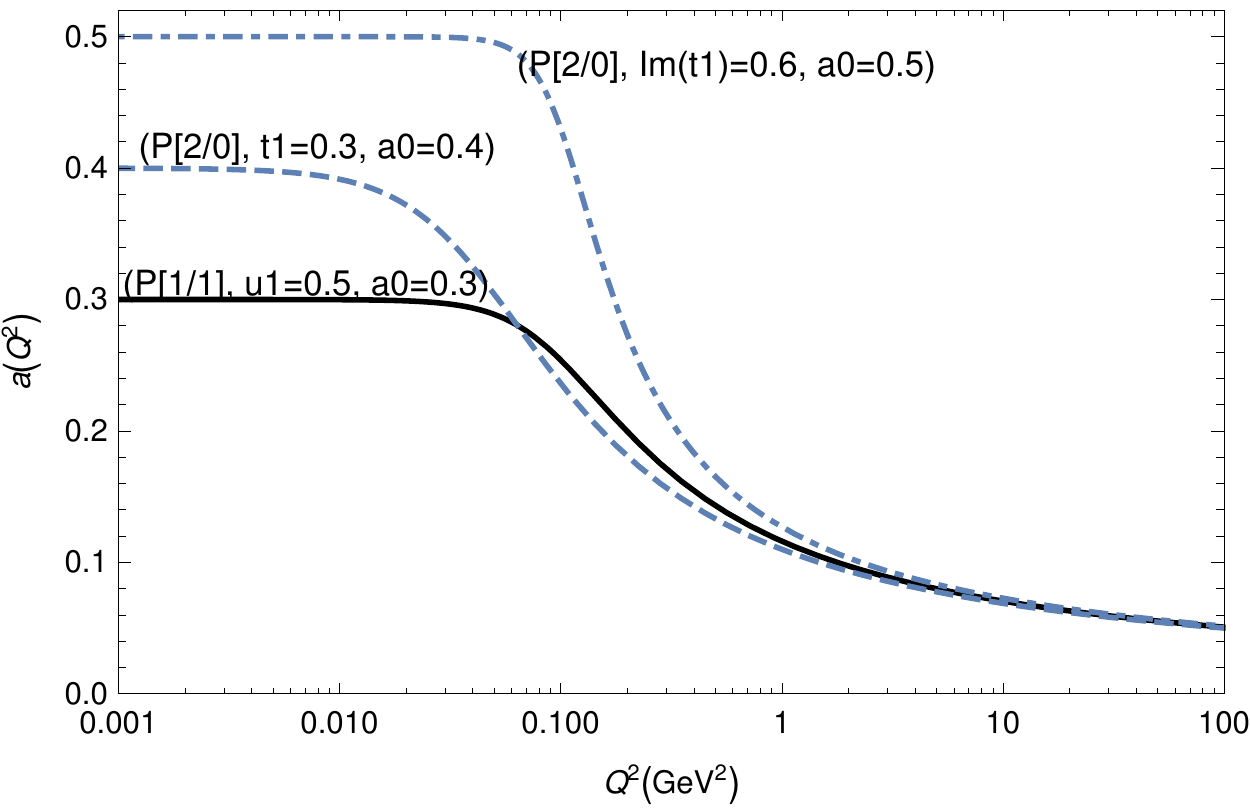}
\vspace{-0.4cm}
\caption{\footnotesize  The running coupling $a(Q^2)$ at spacelike positive $Q^2$, for the three specific cases considered in this Section. The labels $P[M/N]$ refer to the Pad\'e structure of the factor $T_M(Y)/U_N(Y)$ in the $\beta$-function for each case, cf.~Eq.~(\ref{beta1}).}
\label{Figa1all}
\end{figure}

We present in Fig.~\ref{Figa1all} the behaviour of the couplings $a(Q^2)$ for positive $Q^2$ in all three cases considered in this Section. This Figure confirms that the considered class of running couplings has qualitatively similar behaviour in the regime $Q^2>0$, i.e., $a(Q^2)$ is a continuous and monotonically decreasing function of $Q^2$, with finite values in the IR limit at $Q^2=0$.

\begingroup \color{black}
Finally, we present in Figs.~\ref{FigsRhoH}, for the case of the coupling of Sec.~\ref{subs:cr}, the discontinuity (spectral) function $\rho_1(\sigma)={\rm Im} \; a(Q^2=-\sigma - i \epsilon)$ and the corresponding timelike coupling ${\cal H}(s)$ ($s = q^2 \equiv -Q^2 > 0$).
\begin{figure}[htb] 
\begin{minipage}[b]{.49\linewidth}
 \centering\includegraphics[width=88mm]{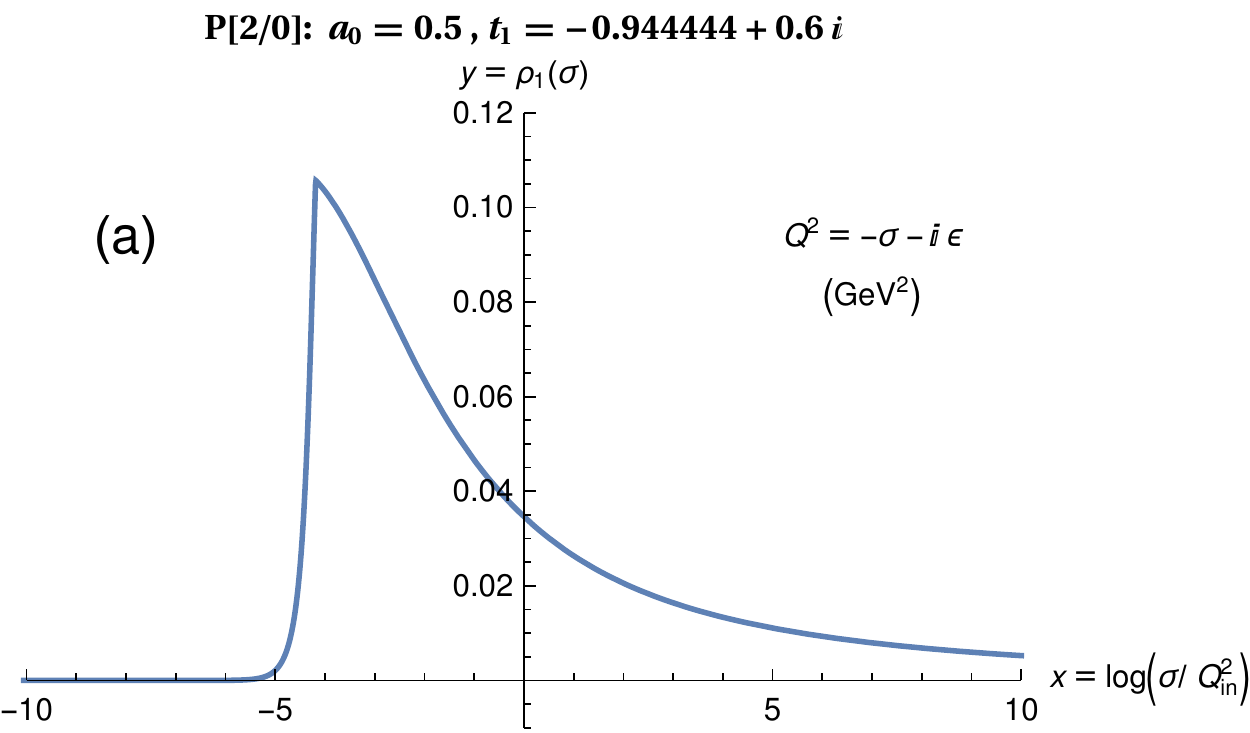}
  \end{minipage}
\begin{minipage}[b]{.49\linewidth}
 \centering\includegraphics[width=88mm]{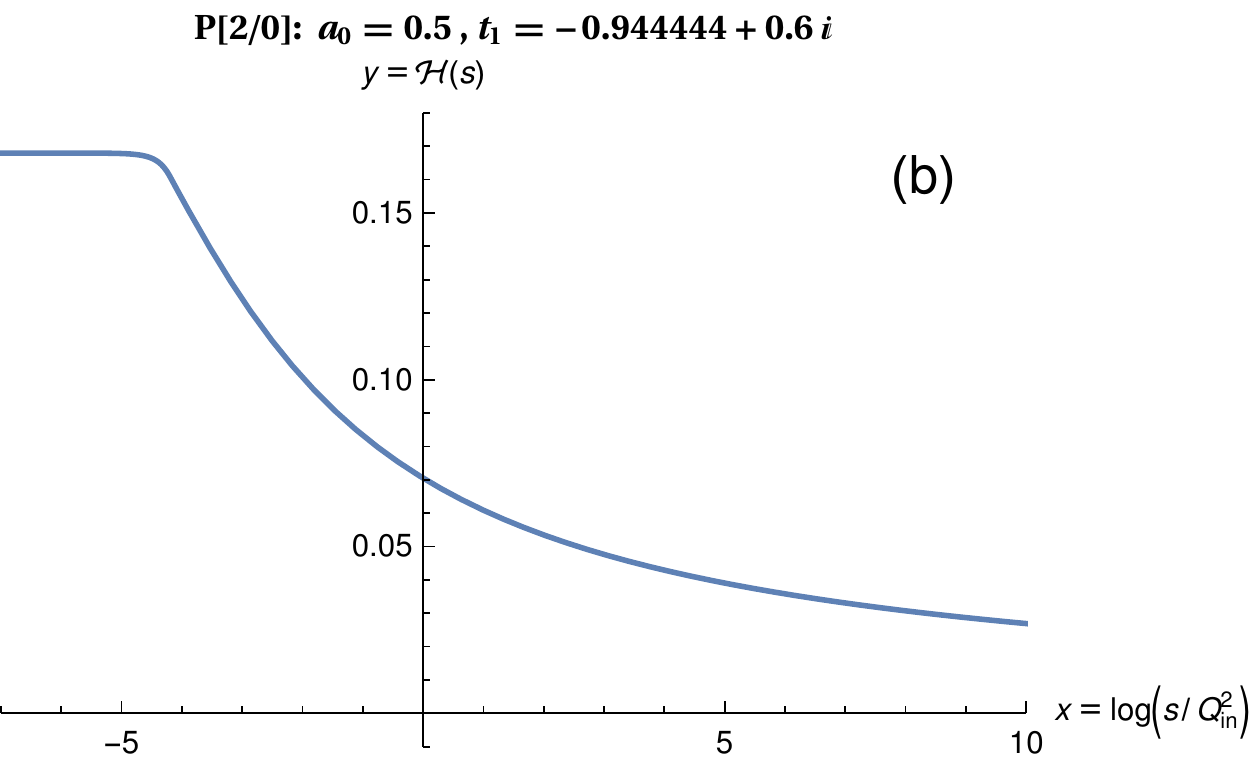}
\end{minipage}
\vspace{-0.2cm}
\caption{\footnotesize  (a) The spectral function  $\rho_1(\sigma)={\rm Im} \; a(Q^2=-\sigma - i \epsilon)$ of the coupling ${\cal H}(s)$, as a function of $\ln(\sigma/Q^2_{\rm in})$; (b) the timelike coupling ${\cal H}(s)$, as a function of $\ln(s/Q^2_{\rm in})$;. We recall that $Q^2_{\rm in}= (2 {\bar m}_c)^2 = 2.54^2 \ {\rm GeV}^2$. }
\label{FigsRhoH}
\end{figure}
The timelike coupling is defined in the usual form \cite{Radyu} (cf.~also \cite{Schre,KP,ShirEPJC})
\be
   {\cal H}(s) = \frac{1}{\pi} \int_s^{+\infty} \frac{d \sigma}{\sigma} \rho_1(\sigma),
\label{calH}
\ee
and fulfills the relation $\pi s d {\cal H}(s)/ds = -\rho_1(s)$.
We notice in Fig.~\ref{FigsRhoH}(b) that ${\cal H}(0) \approx 0.168$ which is less than $a(0) (\equiv a_0) =0.5$; this is a consequence of the Landau singularities of the coupling $a(Q^2)$. Only if $a(Q^2)$ had no Landau singularities, would we obtain  ${\cal H}(0)=a(0)$ \cite{ShirEPJC}.
\endgroup

\section{Summary}
\label{sec:summ}

In this work we presented an algebraic algorithm for finding possible Landau singularities of the pQCD running coupling $a(Q^2)$ in the complex plane of the squared momenta $Q^2$ (first Riemann sheet). We considered a large class of $\beta$-functions, representative of the scenarios where the running coupling $a(Q^2)$ is a monotonic function of $Q^2$ at positive $Q^2$ and ``freezes'' in the IR sector, $a(Q^2) \to a_0$ for $Q^2 \to 0$, where the IR freezing value $a_0$ is considered positive finite. The consideration of the running coupling $a(Q^2) \equiv F(z)$ was performed on the corresponding complex $z$-stripe, $-\pi \leq {\rm Im}(z) < \pi$, where $z=\ln(Q^2/Q^2_{\rm in})$ and $Q^2_{\rm in} > 0$ was an initial scale for the integration of the RGE. The analysis was performed by explicit integration of the RGE which led to the implicit (inverted) solution of the form $z = {\cal H}(F)$. An analysis of this implicit solution than led us to an algebraic procedure for the search of the Landau singularities of $F(z)$ on the $z$-stripe. We considered two types of such singularities, the poles $F(z)=\infty$ and the branching points (for cuts) $\beta(F(z))=\infty$. For illustration, we then presented the mentioned algebraic (and seminumeric) analysis for three specific representative cases of the $\beta$-function, and compared the found Landau singularities with those seen directly by the numerical 2-dimensional integration of the RGE in the entire complex $z$-stripe, the latter approach being numerically demanding. The presented specific cases suggest that our algebraic seminumeric approach is reliable and has high precision in  finding the Landau singularities, while the 2-dimensional integration of the RGE gives these singularities with less precision and sometimes we may miss some of the singular points with this purely numerical method, especially if the numerical scanning over the entire $z$-stripe is made with limited density. Therefore, the presented algebraic seminumeric formalism appears to be useful when we want to find out whether the pQCD running coupling has Landau singularities, and if there are any, to find the location of these singularities with high precision. 

\begin{acknowledgments}
This work was supported in part by the FONDECYT (Chile) Grants No.~1191434 (C.C.), 1180344 (G.C.) and 1181414 (O.O.).  
\end{acknowledgments}

\end{document}